\documentclass[aps,showpacs,twocolumn,
superscriptaddress]{revtex4}
\usepackage{graphicx}% Include figure files
\usepackage{dcolumn}% Align table columns on decimal point
\usepackage{bm}% bold math
\usepackage{color}
\usepackage[normalem]{ulem} % \sout{old text} for strikeout
\usepackage[dvipsnames]{xcolor} % For blue in-text comments
\usepackage{hyperref}
\hypersetup{
  colorlinks=true,        % false: boxed links; true: colored links
  linkcolor=blue,         % color of internal links
  citecolor=cyan,         % color of links to bibliography
}
\usepackage{mathrsfs}
\usepackage[utf8]{inputenc}
\usepackage{mathtools}
\usepackage{doi}
\usepackage{amsmath}
\usepackage{amssymb}
\begin{document}
\title{Charged spinning and magnetized test particles orbiting quantum \\improved charged black holes}

\author{Jose Miguel Ladino}
    \email{jmladinom@unal.edu.co}% Your name
    \affiliation{Universidad Nacional de Colombia, Sede Bogotá. Facultad de Ciencias. Observatorio Astronómico Nacional. Ciudad Universitaria. Bogotá, 111321, Colombia.}
    
\author{Carlos~A.~Benavides-Gallego}
	\email{cabenavidesg@sjtu.edu.cn}
	\affiliation{School of Physics and Astronomy, Shanghai Jiao Tong University, 800 Dongchuan Road, Minhang, Shanghai 200240, PRC.}
    \affiliation{Shanghai Frontiers Science Center of Gravitational Wave Detection, 800 Dongchuan Road, Minhang, Shanghai 200240, People's Republic of China}

\author{Eduard Larra\~{n}aga}
    \email{ealarranga@unal.edu.co}% Your name
    \affiliation{Universidad Nacional de Colombia, Sede Bogotá. Facultad de Ciencias. Observatorio Astronómico Nacional. Ciudad Universitaria. Bogotá, 111321, Colombia.}

\author{Javlon~Rayimbaev}
	\email{javlon@astrin.uz}
	\affiliation{Institute of Fundamental and Applied Research, National Research University TIIAME, Kori Niyoziy 39, Tashkent 100000, Uzbekistan.}
	\affiliation{Akfa University, Kichik Halqa Yuli Street 17,  Tashkent 100095, Uzbekistan.}
	\affiliation{National University of Uzbekistan, Tashkent 100174, Uzbekistan.}
	\affiliation{Tashkent State Technical University, Tashkent 100095, Uzbekistan.}
    
\author{Farrux Abdulxamidov}
	\email{farrux@astrin.uz}
    \affiliation{School of Mathematics and Natural Sciences, New Uzbekistan University, Mustaqillik Ave. 54, Tashkent 100007, Uzbekistan.}
    \affiliation{Ulugh Beg Astronomical Institute, Astronomy St. 33, Tashkent 100052, Uzbekistan.}
    \affiliation{Institute of Nuclear Physics, Ulugbek 1, Tashkent 100214, Uzbekistan.}

\date{\today} % Leave empty to omit a date

\begin{abstract}
    In the present work, we aimed to investigate the dynamics of spinning charged and magnetized test particles around both electrically and magnetically charged quantum-improved black holes. We derive the equations of motion for charged spinning test particles using the Mathisson-Papapetrou-Dixon equations with the Lorentz coupling term. The radius of innermost stable circular orbits (ISCOs), specific angular momentum, and energy for charged spinless, uncharged spinning, and charged spinning test particles around the charged and non-charged quantum-improved black holes are analyzed separately. We found that the quantum parameter increases the maximum spin value, $s_\text{max}$, which leads to the nonphysical motion (superluminal motion) of the charged spinning test particle, whereas the black hole charge decreases its value. We also found that, in contrast to the Reissner Nordstr\"om black hole, spinning charged test particles in the quantum-improved charged black hole have higher $s_\text{max}$; moreover, positively charged spinning particles can have higher values of $s_\text{max}$ near the extreme black hole cases when compared with uncharged spinning particles. Finally, we investigate the magnetized test particle's dynamics around a quantum-improved magnetically charged black hole in Quantum Einstein Gravity using the Hamilton-Jacobi equation. We show that the presence of $\omega$ increases the maximum value of the effective potential and decreases the minimum energy and angular momentum of magnetized particles at their circular orbits. We found an upper constraint in the black hole charge at the ISCO. 
\end{abstract}
%\keywords{first keyword, second keyword, third keyword}
\maketitle
\section{INTRODUCTION} \label{sec:INTRODUCTION}
    General relativity, GR, predicts the formation of singularities after the collapse of massive stars into black holes. However, from the physical point of view, the existence of singularities means that GR breaks down; therefore, a completely new theory of gravity is required to describe the spacetime near them. Nowadays, the scientific community agrees that the quantum effects of gravity play a crucial role in the near region of a singularity, and theorists have deployed a lot of effort into the quest for a quantum theory of gravity, such as the M-theory, string theory, loop quantum gravity, etc.~\cite{DeWitt:1967ub, Aharony:1999ti, Rovelli:1997yv, Ashtekar:2011ni, Thiemann:2001gmi, Barvinsky:1985an, Alvarez:1982zi, Meissner:2004ju, Bojowald:2005epg, Modesto:2011kw, Goroff:1985sz, Cognola:2005de} 

    In traditional approaches to quantum gravity, it is well-known that the Einstein-Hilbert term has been considered a fundamental action~\cite{Reuter:1996cp, Berges:2000ew, Pawlowski:2005xe, Niedermaier:2006wt, Bagnuls:2000ae}. Nevertheless, in contrast with the field theories in flat space (like quantum electrodynamics QED), the Einstein-Hilbert action is non-renormalizable, and a meaningful perturbative analysis becomes difficult~\cite{Reuter:1996cp}. On the other hand, if one assumes that GR is a theory resulting from quantizing a more fundamental theory of gravity, the Einstein-Hilbert term should not be quantized because it becomes an effective action analogous to the Heisenberg-Euler action in QED and, therefore, one should not compare it to the action of electrodynamics~\cite{Reuter:1996cp}. Hence, according to M. Reuter, it is plausible to assume that GR is an effective theory arising from a fundamental one by a ``partial quantization'', which means that Einstein's theory is valid near a non-zero momentum scale $k$, making possible the introduction of a scale-dependent effective action for gravity, from which it is possible to obtain an exact non-perturbative evolution equation governing its renormalization~\cite{Reuter:1996cp}.
    
    In the last decades, there has been an increasing interest in exploring the non-perturbative behavior of quantum gravity~\cite{Reuter:1996cp, Dou:1997fg, Souma:1999at, Lauscher:2001ya, Reuter:2001ag, Lauscher:2001cq, Percacci:2002ie, Percacci:2003jz, Perini:2003wy, Codello:2006in, Litim:2003vp}, and references therein. As mentioned before, in Ref.~\cite{Reuter:1996cp}, the authors proposed a general framework for the treatment of quantum gravity by introducing a scale-dependent action and deriving an exact renormalization group equation; this equation, when applied to the so-called Einstein-Hilbert truncation, allows a non-perturbative approximation to the renormalization group flow of the Newton and cosmological constants. In Refs.~\cite{Dou:1997fg, Percacci:2002ie, Percacci:2003jz, Perini:2003wy}, the authors considered the influence of matter fields. One of the most powerful aspects of the non-perturbative approach of quantum gravity is the existence of a non-Gaussian fixed point for its renormalization group flow. This non-Gaussian fixed point makes quantum gravity a non-perturbatively renormalizable theory, and it plays a fundamental role in the asymptotic safety scenario~\cite{Weinberg:1996kw, Reuter:Book2019}.

    Theoretically, the exact renormalization group flow equation is a powerful tool for finding quantum corrections to solutions of GR,  the Quantum Einstein Gravity, or QEG~\cite{Bonanno:2000ep, Reuter:2010xb, Koch:2013owa, Koch:2014cqa, Bonanno:2001hi, Bonanno:2002zb, Reuter:2003ca, Reuter:2004nv, Ruiz:2021qfp}. In the case of quantum improvement of classical black hole solutions, for example, the Schwarzschild black hole was considered in Ref.~\cite{Bonanno:2000ep},  where the authors investigated the quantum effects in spherically symmetric spacetimes, obtaining the effective quantum spacetime felt by a point-like test mass. The solution is similar to the Reissner Nordstr{\"o}m (RN) black hole, and its conformal structure also depends on its ADM-mass. By computing the Hawking temperature, specific heat capacity, and entropy, the authors conclude that evaporation of the black hole stops when it reaches a critical mass value, $M_\text{cr}$. Furthermore, due to the quantum effects, the quantum spacetime has a smooth de-Sitter core which could be under the cosmic censorship hypothesis~\cite{Bonanno:2000ep}. Hence, the classical singularity $r=0$ can be removed, or it is much milder.
    
    Reference~\cite{Reuter:2010xb} considers the quantum improvement of a rotating black hole, where the horizon structure, the ergo region, the static limit surfaces, and the Penrose process are studied. According to the authors, the quantum corrections become appreciable for lighter black holes. Moreover, in the case of black hole thermodynamics, they found that the first law is modified, and the Bekenstein-Hawking temperature is no longer proportional to the surface gravity. On the other hand, regarding the Penrose process, The authors showed that there exists a minimum mass for the extraction of energy in the improved Kerr spacetime, in contrast with classical black holes, where it is possible to extract energy for arbitrary small mass and angular momentum.

    In Ref.~\cite{Koch:2013owa}, the authors considered quantum corrections to the spherically symmetric Schwarzschild Anti-de Sitter black holes, finding that the cosmological constant play a key role in determining the short-distance structure of quantum-improved black holes. In the asymptotic UV, the solution is universal and similar to the classical Schwarzschild-de Sitter black hole. Therefore, asymptotically safe black holes evaporate completely, and no formation of Planck-size remnants exists.
    
    In Ref~\cite{Ruiz:2021qfp}, O. Ruiz and E. Tuiran investigate the quantum effect in spherically symmetric charged black holes. They found that the horizons are stable except in the extremal case. Moreover, the authors showed the existence of a new extremal condition at the Planck scale that could give clues about the final stage after the evaporation process of the black hole. In contrast to previous results considering axially symmetric spacetimes with null charge, the authors obtained a formula that describes the state function as the sum of the area of the classical event horizon and a quantum correction.
    
    Recently, several works have considered the quantum improvement black hole solutions to investigate their properties~\cite{Mandal:2022stf, Rayimbaev:2020jye, Zuluaga:2021vjc, Atamurotov:2022iwj, Chen:2022dgn, Konoplya:2022hll, Ladino:2022aja}. For example, in Ref.~\cite{Mandal:2022stf}, the authors investigated the geodesic equation for time-like and null-like particles near an improved Schwarzschild black hole. Reference~\cite{Rayimbaev:2020jye} analyzes the dynamics of neutral, electrically charged, and magnetized particles around a renormalized group improved Schwarzschild black hole in the presence of an external asymptotically uniform magnetic field. On the other hand, F.~Zuluaga and L. S\'anchez investigated the quantum effects in the accretion disk around a renormalization group improved Schwarzschild black hole in Ref.~\cite{Zuluaga:2021vjc} and the quantum effects on the black hole shadow and deflection angle in the presence of plasma were studied by F.~Atamurotov et al. in Ref.~\cite{Atamurotov:2022iwj} and Ref.~\cite{Chen:2022dgn} investigates its observational features.

    J.~M.~Ladino and E.~A.~Larra\~naga study the motion of spinning test particles around an improved rotating black hole~\cite{Ladino:2022aja}, where using the Mathisson–Papapetrou–Dixon (MPD) equations and the Tulczyjew spin supplementary condition, the authors investigated the equatorial circular orbits, finding that the event horizon and the radius of the Innermost Stable Circular Orbit (ISCO) for the quantum-improved rotating black hole are smaller than Schwarzschild and Kerr classical solutions. The dynamics of spinning test particles have called the attention of the community, and several works consider different spacetime backgrounds~\cite{Ladino:2022aja, Mikoczi:2019tlu, Yang:2021chw, Benavides-Gallego:2021lqn, Abdulxamidov:2022ofi, Li:2022sjb, Zhang:2022rnn, Ladino:2023laz, Ladino:2023fdq, Plyatsko:2013xza, Hackmann:2014tga, Jefremov:2015gza, Harms:2016ctx, Toshmatov:2019bda, Conde:2019juj, Larranaga:2020ycg, Timogiannis:2021ung, Timogiannis:2022bks}. In this manuscript, we consider the motion of charged spinning test particle motion around quantum-improved charged black holes; we also study the dynamics of magnetized test particles. We organize our paper as follows: in Sec.~\ref{sec:2}, we discuss the characteristics of the quantum-improved charged black hole spacetime. Then, in Sec.~\ref{sec:3}, we introduce the theoretical background to investigate the motion of a charged spinning test particle, which involves the introduction of the modified MPD equations including the force due to the gauge field. In the same section, we also obtain the analytical expression for the effective potential (Sec.\ref{sec:3.2}) used in Sec.~\ref{sec:4} to investigate ISCO, where we consider three cases: charged spinless test particles, uncharged spinning test particles, and charged spinning test particles. Then in Sec.~\ref{sec:5}, we include the case of spinless magnetized particles around a quantum-improved magnetically charged black hole. Finally, in Sec.~\ref{sec:6}, we discuss the results and the most important conclusions. Along the manuscript, we use geometrical units with $G_0=c=1$ and dimensionless variables.

      %I believe leaving the sections in separate files is more organized, change it if you desire 
\section{The Quantum-Improved Charged Black Hole} \label{sec:2}

    In 2000, Bonanno and Reuter presented a new spherically symmetric black hole solution \cite{Bonanno2000} obtained by considering that GR is an effective theory appearing as the low energy limit of a fundamental scheme known as the QEG \cite{QEG01, QEG02, QEG03}. This black hole is similar to the Schwarzschild solution; however, it is characterized by a running gravitational constant, $G(k)$, depending on the energy scale of the theory $k$. Later, a rotating version of this quantum-improved black hole was presented in \cite{Reuter2011, Bambi2013, Torres2017} and more recently, a spherically symmetric charged black hole obtained within the framework of the Einstein-Hilbert truncation in QEG was reported in \cite{Ruiz:2021qfp}. This quantum-improved charged black hole will be the object of study in this paper, and its metric is given by
    \begin{equation}
    ds^2 = - f(r) dt^2 + \frac{dr^2}{f(r)} +r^2 d\Omega^2 \label{eq:BHsolution}
    \end{equation}
    where $d\Omega^2 = d\theta^2 + \sin^2 \theta d\phi^2 $ and the lapse function is
    \begin{equation}
    f(r) = 1 - \frac{2G(r)M}{r} + \frac{G(r) Q^2}{r^2} = \frac{\Delta(r)}{r^2},
    \end{equation}
    with $M$ and $Q$ the mass and electric charge of the black hole. The running gravitational constant can be written, in the limit of long distances related to the Planck length, $r\gg l_p = \sqrt{\frac{\hbar G_0}{c^3}}$, as \cite{Ruiz:2021qfp}
    \begin{equation}
    G(r) = \frac{G_0 r^2}{r^2 + \omega G_0}
    \end{equation}
    where $G_0$ is Newton's gravitational constant and $\omega$ is a parameter, arising from the non-perturbative renormalization group, that measures the quantum effects. In fact, by taking $\omega \rightarrow 0$, the line element recovers the RN solution. Although some studies have restricted the value of this parameter by comparison with the standard perturbative quantization of GR, in this paper, we will consider it as a free positive parameter to describe the general properties of the ISCO for test particles around this black hole.

    The gauge field present in this metric represents a radially symmetric electric field described by the tensor \cite{Ruiz:2021qfp}
    \begin{equation}
    F = dA \label{eq:EMFieldTensor}
    \end{equation}
    with the gauge potential
    \begin{equation}
    A = -\frac{Q}{r} dt. 
    \end{equation}

    The quantum corrected horizons are obtained from the equation
    \begin{equation}
    \Delta (r_\pm) = r_\pm^2 -2G(r_\pm)Mr_\pm + G(r_\pm) Q^2 = 0,
    \end{equation}
    from which one obtains the radii
    \begin{equation}
    r_\pm = G_0M \pm \sqrt{(G_0 M)^2 - G_0\omega - G_0 Q^2 }.
    \label{eq:horizons}
    \end{equation}

    It is straightforward to show that $\omega = 0$ gives the horizons of the RN solution while taking $Q=0$ gives the horizons of the quantum-improved Schwarzschild black hole \cite{Bonanno2000, Torres2017}.

    Additionally, the extreme case of the quantum-improved charged black hole can be easily identified from the discriminant of Eq. \eqref{eq:horizons}, this happens when
    \begin{equation}
    M_\text{ext}=M =\sqrt{\frac{\omega + Q^2 }{G_0} }.
    \label{eq:extreme}
    \end{equation}
%%%%%%%%%%%%%%%%%%% Table %%%%%%%%%%%%%%%%%%
    \begin{table*}[!htb]
    \caption{\label{table:1} ISCO parameters for an unmagnetized, uncharged and spinless particle $(G_0=M=1)$.}
    \begin{tabular}{ccccc}
    \hline$\omega, Q$ \hspace{0.5cm} & Type of black hole  \hspace{0.5cm}  & $r_\text{ISCO}$ & $\hspace{0.5cm}j_\text{ISCO}\hspace{0.5cm}$ & $e_\text{ISCO}$ \\
    \hline$\omega=0, Q=0$ \hspace{0.5cm}  & Schwarzschild  \hspace{0.5cm} & $6$ & $\hspace{0.5cm}3.4641\hspace{0.5cm}$ & $0.9428$ \\
    \hline$\omega=1, Q=0$ \hspace{0.5cm}  & Extreme Improved Schwarzschild  \hspace{0.5cm} & $4.6458$ & $\hspace{0.5cm}3.1765\hspace{0.5cm}$ & $0.9294$ \\
    \hline$0<\omega<1, Q=0$ \hspace{0.5cm}  & Improved Schwarzschild  \hspace{0.5cm} & $(4.6458,6)$ & $\hspace{0.5cm}(3.1765, 3.4641)\hspace{0.5cm}$ & $\left(0.9294, 0.9428\right)$ \\
    \hline$\omega=0, Q=1$ \hspace{0.5cm}  & Extreme RN  \hspace{0.5cm} & $4$ & $\hspace{0.5cm}2.8284\hspace{0.5cm}$ & $0.9186$ \\
    \hline$\omega=0,0<Q<1$ \hspace{0.5cm}  & RN  \hspace{0.5cm}  & $(4,6)$ & $\hspace{0.5cm}(2.8284, 3.4641)\hspace{0.5cm}$ & $\left(0.9186, 0.9428\right)$ \\
    \hline$\omega+Q^2=1$ \hspace{0.5cm}  & Extreme Improved RN  \hspace{0.5cm} & $4.0736$ & $\hspace{0.5cm}2.8676\hspace{0.5cm}$ & $0.9199$ \\
    \hline$0<\omega+Q^2<1$ \hspace{0.5cm}  & Improved RN \hspace{0.5cm}  & $(4.0736,6)$ & $\hspace{0.5cm}(2.8676, 3.4641)\hspace{0.5cm}$ & $\left(0.9199, 0.9428\right) $\\
    \hline
    \end{tabular}
    \end{table*}
%%%%%%%%%%%%%%%%%%%%%%%%%%%%%%%%%%%%%%%%%%%%
    This condition of extremality $M=M_\text{ext}$ is discussed in \cite{Ruiz:2021qfp, Bonanno2000}. The previous expression accurately reproduces the extreme black holes contained within it as particular cases. Specifically, taking $G_0=1$, we obtain the extreme RN solution when $M_\text{ext}=M =Q$, while the extreme quantum-improved Schwarzschild solution is generated when $M_\text{ext}=M =\sqrt{ \omega}$. In any case, for the quantum-improved charged black hole, when $M>M_\text{ext}$, the solutions with two horizons described by Eq. \eqref{eq:horizons} are obtained. Taking $G_0=M=1$ and since $M>M_\text{ext}$, $\omega$ takes values in the range $
   0< \omega \leq 1-Q^2$.

\begin{figure}[!htb]
    \centering
    \includegraphics[width=0.99\linewidth]{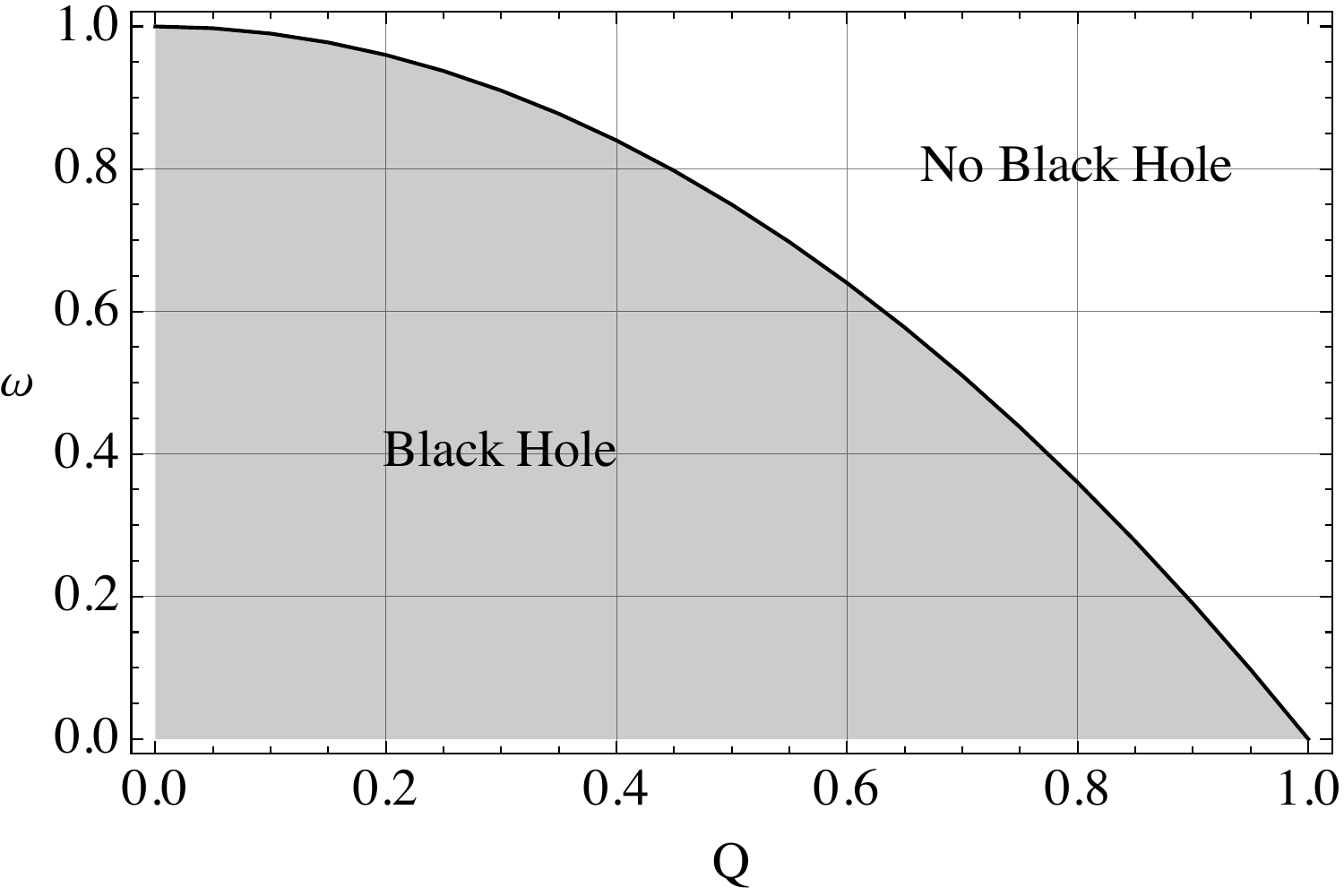}
    \caption{Relation between $Q$ and $\omega$ parameters of the black hole. \label{horizonbhp}}
    \end{figure}

    When $Q \rightarrow 1$, the parameter $\omega \rightarrow 0$, this shows that the parameter $\omega$ must satisfy certain conditions for the space-time to be globally hyperbolic. On the other hand, when $M<M_\text{ext}$, the space-time contains a naked singularity. This violates the conditions for a globally hyperbolic space-time and is therefore not considered in our analysis. In Fig.~\ref{horizonbhp}, the relation between $\omega$, and $Q$ is visualized. It is possible to see two regions. The gray region represents the allowable parameter values, and the white one corresponds to values that make the black hole a naked singularity.

    %%%%%%%%%%%%%%%%%%% Figure I %%%%%%%%%%%%%%%%%%%%%%
    %%%%%%%%%%%%%%%%%%%%%%%%%%%%%%%%%%%%%%%%%%%%%%%%%%%
    
    To help facilitate comparison, we have compiled a list of the ISCO parameters, the radius $r_\text{ISCO}$, the energy $e_\text{ISCO}$ and the orbital angular momentum $l_\text{ISCO}$,  for an uncharged and spinless particle in Table \ref{table:1}, including quantum-improved black holes, as well as their particular and extreme cases.

\section{Equations of Motion for a Charged Spinning Test Particle} \label{sec:3}

    %%%%%%%%%%%%%%%%%%% Figure II %%%%%%%%%%%%%%%%%%%
    \begin{figure*}[!htb]
    \centering
    \includegraphics[width=1\linewidth]{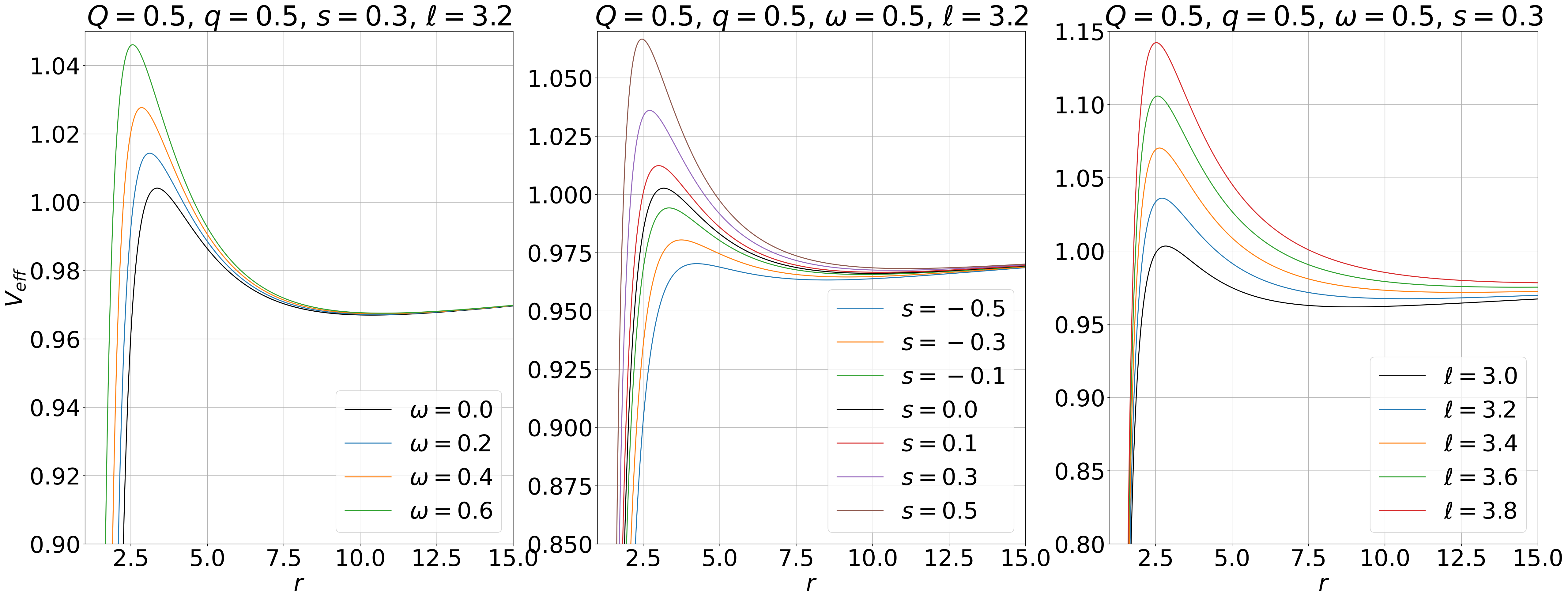}
    \caption{Radial dependence of the effective potential for different values of $\omega$, $s$ and $\ell$; left, center and right panels, respectively.}
    \label{veffp}
    \end{figure*}
    %%%%%%%%%%%%%%%%%%%%%%%%%%%%%%%%%%%%%%%%%%%%%%%%%%

    The equations of motion for a charged  spinning test particle around a charged black hole are given by a modification of the MPD equations including the force due to the gauge field. Then, the system reads
    \begin{align}
        \frac{D p^\mu}{d\lambda} = &-\frac{1}{2} R^\mu_{\nu \rho \sigma} v^\nu S^{\rho \sigma} - m q F^\mu _\nu v^\nu \label{eq:MPD1}\\
        \frac{D S^{\mu \nu}}{d\lambda} =& p^\mu v^\nu - v^\mu p^\nu, \label{eq:MPD2}
    \end{align}
    where the velocity vector of the test particle is represented by $v^\mu$, its momentum by $p^\mu$, its electric charge is $q$ and $m$ corresponds to the dynamical rest mass defined by
    \begin{equation}
    m^2 = - p_\mu p ^\mu.
    \end{equation}

    In the MPD equations, we also have the Riemann tensor of the spacetime, $R^\mu_{\nu \rho \sigma}$, the absolute derivative  $\frac{D}{d\lambda} = u^\mu \nabla_\mu$,  the electromagnetic field tensor, $F_{\mu \nu}$, defined in Eq. (\ref{eq:EMFieldTensor}) and the spin tensor, $S^{\mu \nu} = - S^{\nu \mu}$,  which defines the particle's spin as
    \begin{equation}
    s^2 = \frac{1}{2} S_{\mu \nu} S^{\mu \nu}.
    \end{equation}
    In order to solve this system of equations, it is possible to include a constraint involving the spin tensor. In this work, we consider the well-known Tulczyjew spin supplementary condition, which reads
    \begin{equation}
    p_\mu S^{\mu \nu}  = 0.
    \end{equation}

%%%%%%%%%%%%%%%%%% Subsection %%%%%%%%%%%%%
\subsection{Conserved Quantities}

    Along the motion of the charged spinning test particle, we can identify a conserved quantity $C_k$ related to the existence of a Killing vector field $k$ as
    \begin{equation}
    C_k = p^\mu k_\mu + \frac{1}{2} S^{\mu \nu} \Delta_\mu k_\nu + mq k^\mu A_\mu.
    \end{equation}
    The quantum-improved black hole spacetime (\ref{eq:BHsolution}) admits the existence of two Killing vectors. The first one, $\xi = \frac{\partial}{\partial t}$, is related to the conservation of energy per unit mass, 
    \begin{equation}
    e = - u^\mu k_\mu - \frac{1}{2m} S^{\mu \nu} \nabla_\mu \xi_\nu - q \Phi,
    \end{equation}
    where we introduce the normalized momentum $u^\mu = {p^\mu}/{m}$ and use the non-vanishing component of $A_\mu$, which is identified with the scalar potential $\Phi = -{Q}/{r}$. The second Killing vector, $\varphi = \frac{\partial}{\partial \phi}$, is related to the conservation of the angular momentum per unit mass,
    \begin{equation}
    j = - u^\mu \varphi_\mu - \frac{1}{2m} S^{\mu \nu} \nabla_\mu \varphi_\nu.
    \end{equation}
    
    In order to calculate these quantities, we introduce a local orthonormal tetrad field and its inverse, 
    \begin{align}
    e_\mu^{(t)} dx^\mu=& \frac{\sqrt{\Delta}}{r} dt; 
    & e^\mu_{(t)} \partial_\mu=& \frac{r}{\sqrt{\Delta}}\partial_t\\
    e_\mu^{(r)} dx^\mu=& \frac{r}{\sqrt{\Delta}} dr; 
    & e^\mu_{(r)} \partial_\mu= &\frac{\sqrt{\Delta}}{r}\partial_r\\
    e_\mu^{(\theta)} dx^\mu=&  rd\theta; 
    & e^\mu_{(\theta)} \partial_\mu= &\frac{1}{r}\partial_\theta\\
    e_\mu^{(\phi)} dx^\mu =&  r\sin \theta d\phi; 
    & e^\mu_{(\phi)} \partial_\mu = &\frac{1}{r\sin \theta} \partial_\phi.
    \end{align}
    Since we focus our work on equatorial circular motion, we set $\theta= \frac{\pi}{2}$ and we can simplify the analysis by assuming that, in the tetrad frame, the only non-vanishing component of the spin vector is $s^{(\theta )}= -s $. Similarly, the momentum vector will have
    \begin{equation}
    p^{(\theta)} = 0
    \end{equation}
    and from the relation between the spin vector and the spin tensor,
    \begin{equation}
        S^{(\alpha) (\beta)} = \epsilon^{(\alpha) (\beta)}_{\quad(\rho)(\sigma)} p^{(\rho)} s^{(\sigma)},
    \end{equation}
    we obtain the non-zero components
    \begin{align}
        S^{(t)(r)} = &-sp^{(\phi)}\\
        S^{(t)(\phi)} = & sp^{(r)}\\
        S^{(r)(\phi)} = & sp^{(t)}.
    \end{align}
    Using these results, the conserved quantities are expressed in the tetrad frame as
    \begin{align}
        e = & \frac{\sqrt{\Delta}}{r} u^{(t)} + \frac{G_0 [r(Mr-Q^2) - M\omega G_0]}{(r^2 + \omega G_0)^2}s u^{(\phi)} + \frac{qQ}{r}\label{eq:energy}\\
        j = &r u^{(\phi)} + \frac{\sqrt{\Delta}}{r}s u^{(t)}.\label{eq:angularmomentum}
    \end{align}
    Hence, with the previous results, we can now calculate the momentum components governing the motion of the charged spinning test particles.
%%%%%%%%%%%%%%%%%% Subsection %%%%%%%%%%%%%%%%%%%%%%%%%
\subsection{The Effective Potential}\label{sec:3.2}

    To define the effective potential for charged spinning test particles moving in the equatorial plane, we first use Eqs.~(\ref{eq:energy}) and  (\ref{eq:angularmomentum}) to obtain the components of the normalized momentum in the tetrad frame,
    \begin{align}
        u^{(t)} = &\frac{r X}{\sqrt{\Delta} Z}\\
        u^{(\phi)} = & \frac{Y}{Z}
    \end{align}
    where
    \begin{align}
        X = & W_1 (re-qQ) + W_2 j\\
        Y = &W_1 \left(-se + j + \frac{qQ}{r}s \right)\\
        Z = &W_1r + W_2 s\\
        W_1 = &(r^2 + \omega G_0)^2  \\
        W_2 = &-G_0[r(Mr-Q^2) - M\omega G_0]s.
    \end{align}
    The remaining component is obtaining using the normalization condition $u_{(\alpha)} u^{(\alpha)} = -1$, giving by 
    \begin{equation}
        u^{(r)} = \pm \frac{\sqrt{R}}{\sqrt{\Delta} Z}, \label{eq:ucomponentr}
    \end{equation}
    with
    \begin{equation}
        R = r^2 X^2 - Y^2 \Delta - Z^2 \Delta.
    \end{equation}
    The signs in Eq.~(\ref{eq:ucomponentr}) represent radially outgoing $(+)$ or ingoing $(-)$ particles. Then, with the help of the inverse tetrad, it is possible to write the components of the normalized momentum in the general frame as    
    \begin{align}
        u^{t} = &\frac{r^2 X}{\Delta Z}\\
        u^{r} = &\pm \frac{\sqrt{R}}{r Z} \\
        u^{\phi} = & \frac{Y}{rZ}.
    \end{align}
    
    The condition for circular motion may be imposed as $(u^r)^2 = 0$, which evidently implies that
    \begin{eqnarray}
    \label{eq:circularmotioncondition}
    \nonumber 
        R &=& Ae^2 + B + C 
        \\ \nonumber 
        &=& A\left( e- \frac{-B + \sqrt{B^2 - 4AC}}{2A}\right)
        \\ &\times& \left( e+ \frac{-B + \sqrt{B^2 - 4AC}}{2A}\right) = 0, 
    \end{eqnarray}    
    where we have defined
    \begin{align}
        A = &W_1^2 \left(r^4  - \Delta  s^2 \right) \\
        B = & 2 W_1(W_2 j - W_1 qQ)  r^3 + 2 W_1^2 \Delta \left(j + \frac{qQ}{r} s\right) s \\
        C = & (W_2 j - W_1 qQ)^2 r^2 - W_1^2 \Delta \left( j + \frac{qQ}{r}s \right)^2  - Z^2 \Delta.
    \end{align}
    The roots of Eq.~(\ref{eq:circularmotioncondition}) define the effective potential as
    \begin{equation}
        V_\text{eff} (r) = \frac{-B + \sqrt{B^2 - 4AC}}{2A}.
        \label{potentialgeneral}
    \end{equation}
    
    Figure \ref{veffp} shows the radial dependence of the effective potential for different values of  $\omega$, $s$, and $\ell$ of the particle while keeping its charge, $q$, and the black hole's charge, $Q$, fixed. The figure consists of three panels. First, we explore the effect of the parameter $\omega$ on the effective potential in the left panel, where it is possible to see that an increment in $\omega$ increases the effective potential. In the figure, we set $Q$ and $q$ to $0.5$, the particle's spin is $s = 0.3$, and its orbital angular momentum is $\ell= 3.2$. In the middle panel, we study the effect of $s$ on the effective potential. From the figure, it is possible to see a similar trend as in the previous case $\omega$, namely, an increment in the particle spin causes an increment in $V_\text{eff}$. The charge, orbital angular momentum, and $\omega$ were not changed, with $\omega$ set to 0.5. Finally, the right panel shows the effect of $\ell$ on $V_\text{eff}$. Here,  as with the other parameters, an increase in the orbital angular momentum leads to an increment in the effective potential. 
    
    %%%%%%%%%%%%%%%%%%%% Fig 3 %%%%%%%%%%%%%%%%%%%%%%%%
    \begin{figure}[ht!]\centering
    \includegraphics[width=0.85\linewidth]{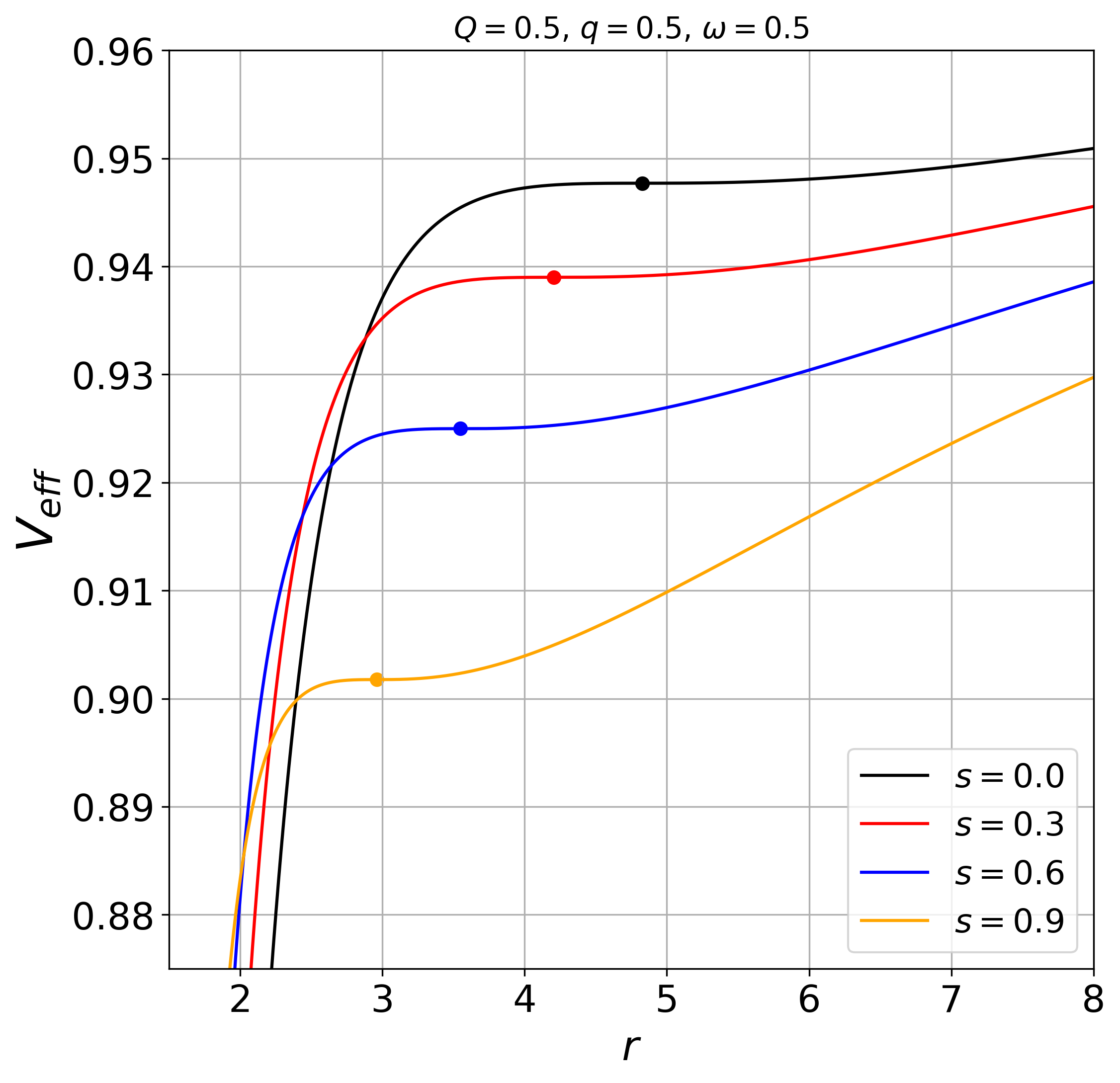}
    \includegraphics[width=0.85\linewidth]{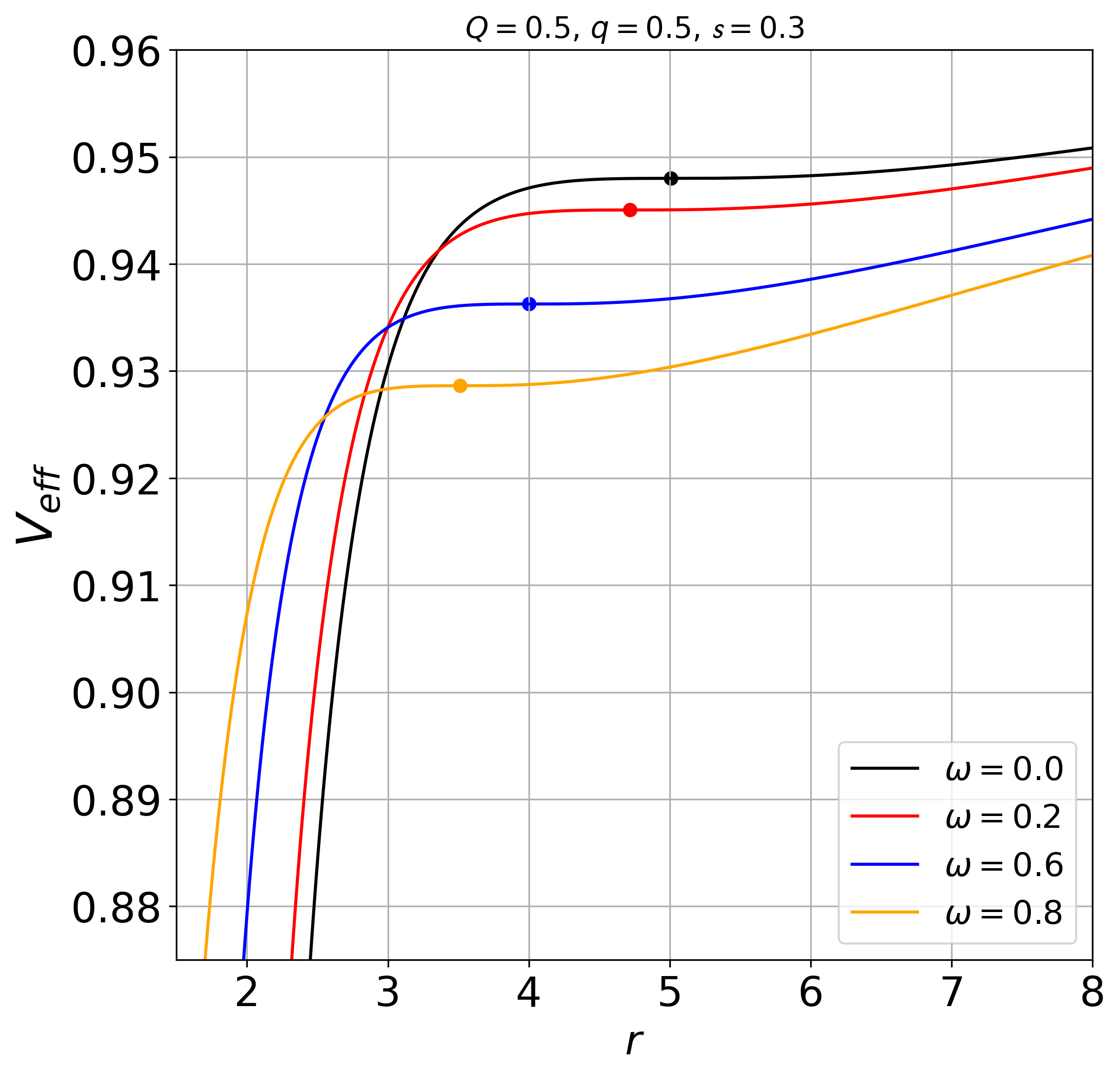}
     \caption{Radial dependence of the effective potential with ISCO points for the different values of $s$ spin of the particle (top panel), and $\omega$ parameter (bottom panel).}\label{veffwi}
     \end{figure}
    %%%%%%%%%%%%%%%%%%%%%%%%%%%%%%%%%%%%%%%%%%%%%%%%%%%

     %%%%%%%%%%%%%%%%%%%% Fig 4 %%%%%%%%%%%%%%%%%%%%%%%%

\begin{figure*}[!htb]
    \centering
\includegraphics[width=0.85\linewidth]{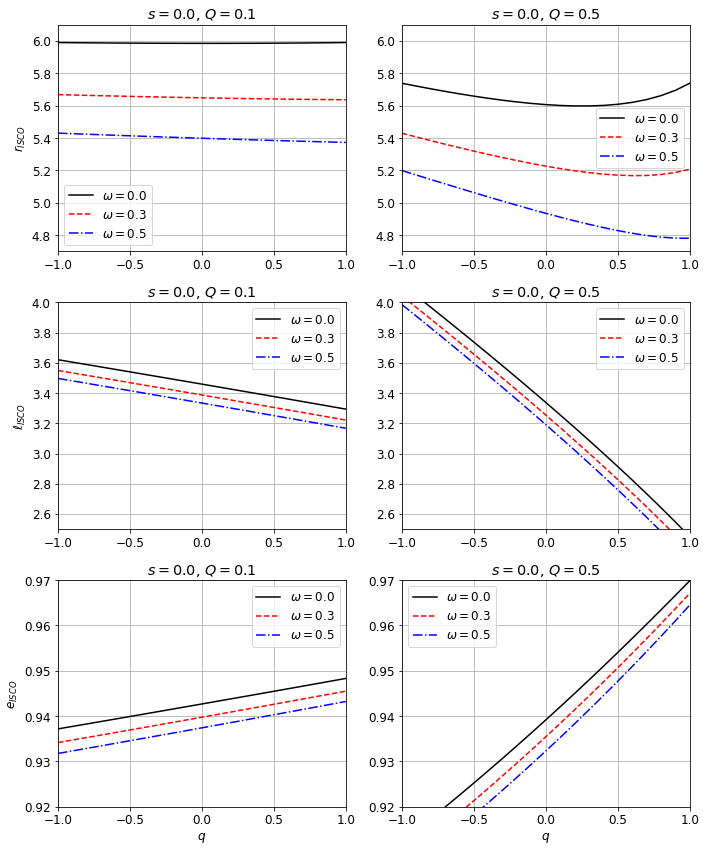}
    \caption{ISCO parameters as a  function of the electric charge $q$ of a spinless test particle. All plots use $G_0 = M = 1$. }
    \label{fig:spinless}
\end{figure*}
%%%%%%%%%%%%%%%%%%%%%%%%%%%%%%%%%%%%%%%%%%%%
    
    Figure~\ref{veffwi} shows the radial dependence of the effective potential with some ISCO points. In the top panel, the spin of the particle varies, with $s = 0, \ell_{ISCO} =2.76 $  (black), $s = 0.3, \ell_{ISCO} =2.54 $  (red), $s = 0.6, \ell_{ISCO} =2.25 $  (blue), $s = 0.9, \ell_{ISCO} =1.87 $  (orange). It can be seen that an increment in $s$ produces a decrease in the ISCO radius and the orbital angular momentum at that radius. Therefore, spinning test particles can move closer to the black hole's center. A similar trend can be seen in the bottom panel, where $\omega$ changes: $\omega = 0, \ell_{ISCO} =2.72 $  (black), $\omega = 0.2, \ell_{ISCO} =2.66 $  (red), $\omega = 0.6, \ell_{ISCO} =2.49 $  (blue), $\omega = 0.8, \ell_{ISCO} =2.37 $ (orange).

    Additionally, it is worth noting that when $s$ is too large, the four-velocity $v^\mu$ may not be timelike. To address this issue, a superluminal constraint is imposed, as was done in \cite{Zhang2018}. This constraint for  equatorial circular motion is represented by
    \begin{equation}
        \frac{v^\mu v_\mu}{(v^t)^2} = -f(r) + \frac{\dot{r}^2}{f(r)}  + r^2 \dot{\phi}^2 < 0,
    \label{eq:superluminal}
    \end{equation}
    which limits the values of spin $s$ that lead to the nonphysical motion of the charged spinning test particle. This superluminal constraint defines a minimum and maximum value for $s$; in this paper, we focus on the superluminal bound denoted as $s_\text{max}$. 

    We can also introduce the orbital angular momentum  $\ell$, with the total angular momentum defined as $j=\ell+s$. Therefore, we can determine the ISCO parameters by numerically solving a system of equations involving the conditions 
    \begin{equation} \label{condition}
        V_\text{eff}=e\, \quad  \frac{dV_\text{eff}}{dr} = 0,\quad
        \frac{d^2V_\text{eff}}{dr^2} = 0.
    \end{equation}
    Solving this system of three equations, we can obtain the ISCO parameters $r_\text{ISCO}$, $e_\text{ISCO}$ and $\ell_\text{ISCO}$ for different values of $s$, $q$ $Q$ and $ \omega $. In Table~\ref{table:1}, we share the simplest results, where the possible values for the ISCO parameters for uncharged ($q=0$)  and spinless ($s=0$) particles are shown. 
    
    In the next section, we investigate test particles with electric charge $q$, spin $s$, and with both parameters considered simultaneously.

%%%%%%%%%%%%%%%%%%%% Fig 5 %%%%%%%%%%%%%%%%%%%%%%%%

    \begin{figure*}[!htb]
    \centering
    \includegraphics[width=1\linewidth]{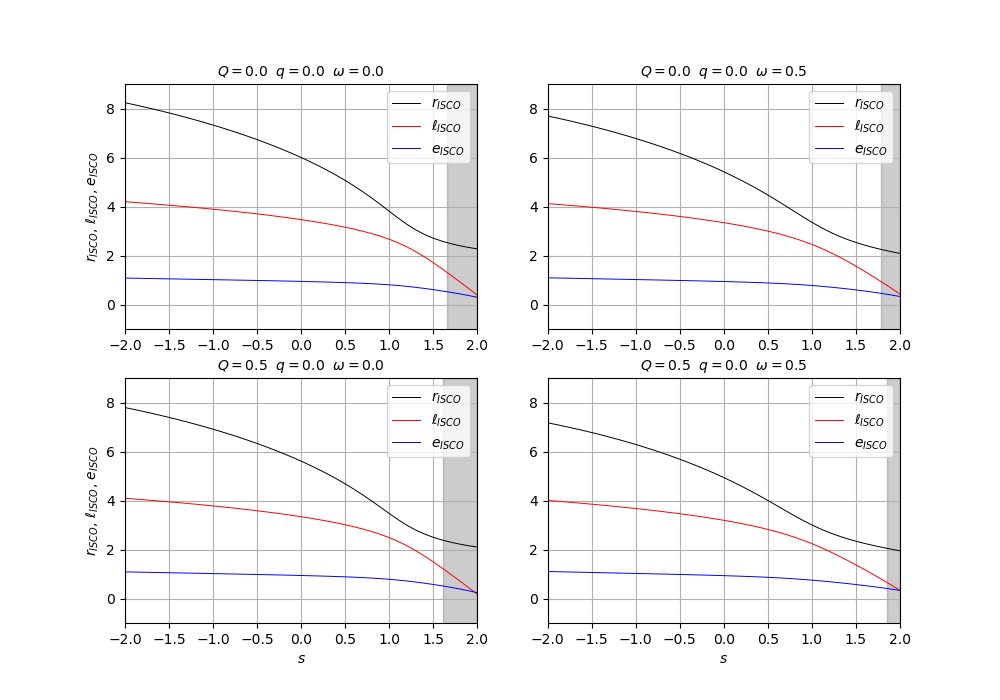}
    \caption{ISCO parameters as a function of the spin $s$ of an uncharged test particle moving in some particular cases: the Schwarzschild spacetime (up/left), the quantum-improved Schwarzschild background with $\omega = 0.5$ (up/right), the RN solution (down/left) and the quantum-improved charged black hole background with $\omega = 0.5$. The shaded region indicates nonphysical motion according to the superluminal condition. (\ref{eq:superluminal}). All plots use $G_0 = M = 1$. }
    \label{fig:fig1}
    \end{figure*}

%%%%%%%%%%%%%%%%%%%%%%%%%%%%%%%%%%%%%%%%%%%%%%%%%%

\section{ISCO Parameters of Charged Spinning Test Particles} \label{sec:4}
%%%%%%%%%%%%%%%%%%%% Fig 6 %%%%%%%%%%%%%%%%%%%%%%%%

    \begin{figure*}[!htb]
    \centering
    \includegraphics[width=1\linewidth]{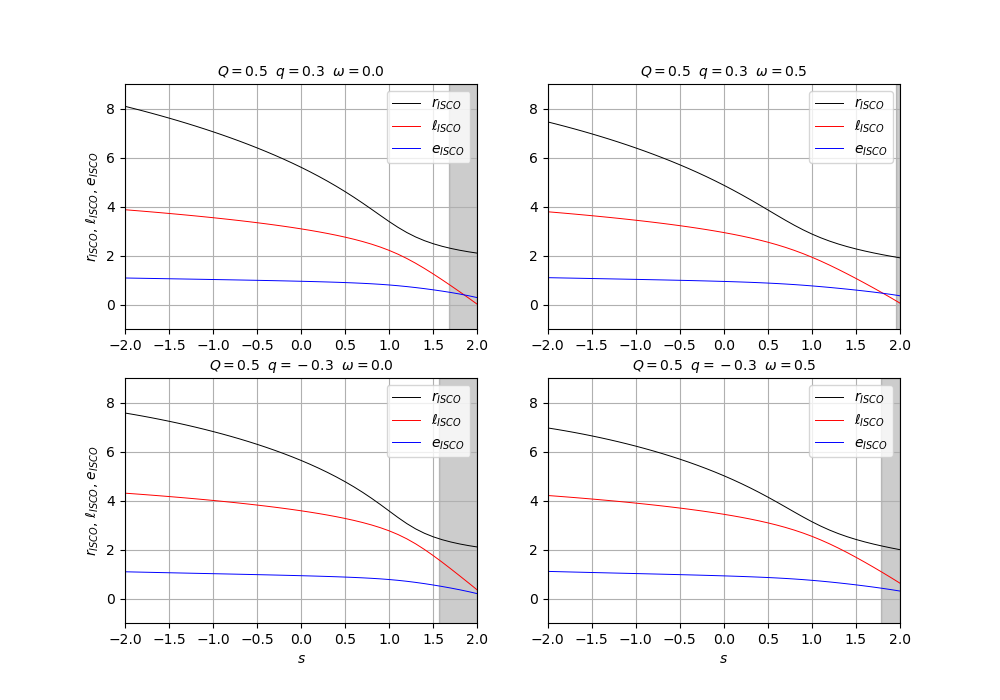}
    \caption{ISCO parameters as a function of the spin $s$ of an electrically charged test particle moving around the quantum-improved charged black hole. Note the dependence on the charge of the test particle, $q$, and on the parameter $\omega$. The shaded region indicates nonphysical motion according to the superluminal condition. (\ref{eq:superluminal}). All plots use $G_0 = M = 1$. }
    \label{fig:fig2}
    \end{figure*}
%%%%%%%%%%%%%%%%%%%%%%%%%%%%%%%%%%%%%%%%%%%%%%%%%%%
%%%%%%%%%%%%%%%%%%%% Fig 7 %%%%%%%%%%%%%%%%%%%%%%%%

    \begin{figure}[!htb]
    \centering
    \includegraphics[width=1\linewidth ]{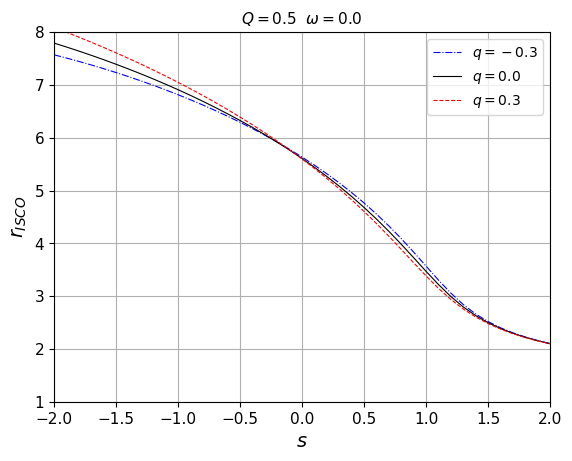}
    \includegraphics[width=1\linewidth]{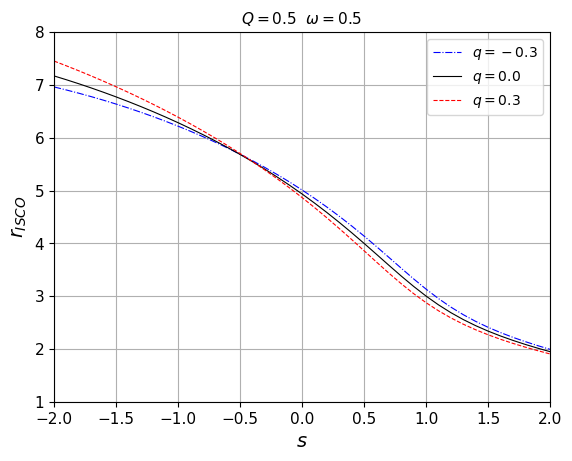}
    \caption{The radius of the ISCO  as a function of the spin $s$ of a charged test particle moving around the quantum-improved charged black hole. The figure shows the behavior for different values of the electric charge of the test particle. All plots use $G_0 = M = 1$.}
    \label{fig:fig9}
    \end{figure}
%%%%%%%%%%%%%%%%%%%%%%%%%%%%%%%%%%%%%%%%%%%%%%%%%%%
%%%%%%%%%%%%%%%%%%%% Fig 8 %%%%%%%%%%%%%%%%%%%%%%%%
    \begin{figure}[!htb]
    \centering
    \includegraphics[width=1\linewidth]{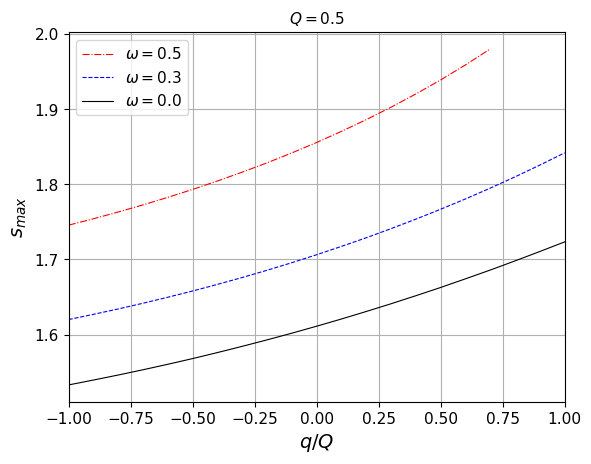}
    \caption{Maximum spin $s_{max}$ allowed by the superluminal condition (\ref{eq:superluminal}) for a charged test particle moving at the ISCO as a function of the charge ratio $q/Q$. All plots use $G_0 = M = 1$. }
    \label{fig:fig11}
    \end{figure}
%%%%%%%%%%%%%%%%%%%%%%%%%%%%%%%%%%%%%%%%%%%%

\subsection{ISCO of charged spinless test particles} \label{sec:5.1}

    If we assume the simple case of motion of charged spinless ($s=0$) test particles, we can only study the dynamic effects produced on the ISCO parameters by the electric charge $q$ of the test particle. Taking $s=0$, the effective potential of Eq.~\eqref{potentialgeneral} reduces to
    \begin{equation}
    V_\text{eff} (r) = \frac{qQ}{r}+\frac{\sqrt{\Delta (j^2+r^2)}}{r^2}.
    \end{equation}
    In Fig. \ref{fig:spinless}, we present the behavior of the ISCO parameters as a function of the electric charge $q$ for a spinless test particle orbiting different black hole solutions. Note that the black curve corresponds to the RN spacetime ($\omega=0.0$), while the red and blue curves to the quantum-improved charged black hole solutions; $\omega=0.3$ and $\omega=0.5$, respectively. In the figure, the behavior for $Q=0.1$ is depicted in the left panels, and the right panels show the behavior when $Q=0.5$. Here we have excluded the Schwarzschild and the quantum-improved Schwarzschild metric cases, as there is no interaction between $q$ and the geometric backgrounds of these solutions.

    In Fig. \ref{fig:spinless} is evident that as the quantum parameter $\omega$ increases, the values of the ISCO parameters $r_\text{ISCO}$, $e_\text{ISCO}$, and $\ell_\text{ISCO}$ decrease, regardless of the electric charge $q$. Also, when $\omega=0$, $r_\text{ISCO}$ exhibits a parabolic behavior in response to $q$ and $Q$. As $\omega$ increases, the decreasing trend of $r_\text{ISCO}$ as a function of $q$ becomes more prominent, especially for small values of $Q$. Hence, we can conclude that the quantum correction, represented by $\omega$, alters the parabolic behavior of $r_\text{ISCO}$ with respect to $q$ (a characteristic of the RN metric) and instead causes it to exhibit a decreasing monotonic trend. Furthermore, for a given increase in $Q$, $r_\text{ISCO}$, $\ell_\text{ISCO}$ and $e_\text{ISCO}$ becomes smaller. Generally, larger values of $q$ lead to smaller values of $\ell_\text{ISCO}$ and larger values of $e_\text{ISCO}$. These corrections on the ISCO parameters due to $q$ and $Q$ agree with the results reported in \cite{Zhang2019, Pugliese:2011py} for charged spinless test particles around RN and Kerr-Newman spacetimes. 

\subsection{ISCO of uncharged spinning test particles} \label{sec:5.2}

    Setting $q=0$, we can reduce the general description of the effective potential of Eq.~\eqref{potentialgeneral} to analyze the behavior of the ISCO parameters of uncharged spinning test particles.

    In Fig.~\ref{fig:fig1}, we show the behavior of the ISCO parameters as a function of the spin $s$ of an uncharged test particle moving around the quantum-improved charged black hole background and their particular cases. The effects resulting from the spin $s$ are typical and well-known \cite{Ladino:2022aja, Yang:2021chw, Ladino:2023laz, Toshmatov:2019bda, Conde:2019juj, Larranaga:2020ycg, Zhang2018}. It is clear from the figure that the ISCO parameters for the quantum-improved charged black hole are smaller than those of their respective particular cases. Although the corrections resulting from the electric charge $Q$ and the quantum parameter $\omega$ are similar, the former is less perceptible. Therefore, we can conclude that the ISCO parameters for the quantum-improved Schwarzschild black hole are smaller than those for the RN black hole. However, the largest ISCO properties are obtained for the Schwarzschild black hole.

    In Fig.~\ref{fig:fig1}, we compare the superluminal motion resulting from the spin $s$ for the different metrics under study. The figure clearly shows that test particles orbiting the quantum-improved charged black hole reach the nonphysical motion at larger spin values. Interestingly, for the parameter values chosen, we see that spinning test particles around the RN black hole reach the superluminal bound faster than those around the Schwarzschild black hole, followed by the quantum-improved black hole cases. Therefore, the quantum parameter $\omega$ increases the maximum spin $s_\text{max}$, while the electric charge $Q$ decreases it. When considering both parameters simultaneously, $s_\text{max}$ becomes much smaller than the case in which only $\omega$ is taken into account.

\subsection{ISCO of charged spinning test particles} \label{sec:5.3}

    We now consider the general case of circular equatorial motion of charged spinning test particles around metrics backgrounds with electric charge $Q$, namely the RN black hole (with $\omega=0$) and the quantum-improved charged black hole (with $\omega \neq 0$). To achieve this, we will use the effective potential derived in Eq.~\eqref{potentialgeneral} while taking into account both $q \neq 0$ and $s \neq 0$ simultaneously. 

    In Figs. \ref{fig:fig2} and \ref{fig:fig9}, we illustrate how the ISCO parameters vary with the spin $s$ of a charged spinning test particle orbiting around the RN and quantum-improved charged black holes. Figure~\ref{fig:fig2} illustrates the typical behavior of the ISCO parameters as a function of the spin $s$. Here, we observe that irrespective of the electric charge $q$ of the test particle, the ISCO parameters decrease with an increase in $s$ until the shaded region is reached, indicating the limit beyond which particles attain superluminal speeds. Furthermore, we note that an increase in the quantum parameter $\omega$ leads to a decrease in the ISCO parameters $r_\text{ISCO}$, $\ell_\text{ISCO}$, and $e_{ISCO}$, regardless of the value of $q$. Additionally, independent of the value of spin $s$, it can be seen that higher values of electric charge $q$ lead to lower values of $\ell_\text{ISCO}$ and higher values of $e_\text{ISCO}$.

    The impact of the particle's charge $q$ on its ISCO radius $r_\text{ISCO}$ is somewhat intricate to analyze, as its effect varies depending on the spin $s$. Nevertheless, Fig.~\ref{fig:fig9} provides a clear illustration of this behavior. There, we show the corrections resulting from positive and negative values of $q$, indicating that simultaneous changes in the particle's charge and spin do not produce monotonic effects on the ISCO. For a fixed $q$, $r_\text{ISCO}$ decreases monotonically with $s$, but for a given value of $s$, this is not the case always. Nonetheless, the plot reinforces the fact that $r_\text{ISCO}$ is smaller for a charged spinning test particle orbiting around the quantum-improved charged black hole than around the RN black hole, regardless of the values of $q$ and $s$. These modifications made to the ISCO parameters for charged spinning particles, when taking $q$ and $s$ simultaneously, are consistent with the results reported in \cite{Zhang2019} for RN black holes.

    Figure.~\ref{fig:fig9} also shows that variations in the electric charge $q$ of the test particle significantly affect the maximum spin $s_\text{max}$ that delimits the region of superluminal motion. The impact of $q$ on $s_\text{max}$ varies for each black hole solution. Nevertheless, the quantum-improved charged black hole yields a higher maximum spin value $s_\text{max}$ than the RN spacetime, regardless of the value of $q$. Furthermore, $s_\text{max}$ can attain larger values as $q$ increases, so corrections due to negative $q$ values reduce $s_\text{max}$.

    The corrections due to the electric charge $q$ on the maximum spin $s_\text{max}$ can be better understood by examining Fig.~\ref{fig:fig11}, which illustrates the behavior of $s_\text{max}$ allowed by the superluminal condition as a function of the charge ratio $q/Q$. Here, we confirm that higher values of the quantum parameter $\omega$ correspond to higher values of $s_\text{max}$. Moreover, larger values of $q$ result in higher values of $s_\text{max}$. Thus, a charged spinning test particle can orbit with higher spin values $s$ if it possesses a more positive electrical charge $q$ or larger geometric parameters (without exceeding the extreme case of the solution).

    The physically possible maximum value of spin $s_\text{max}$ of uncharged test particles is reached in the extreme case of the black hole solution, where the geometric parameters of the spacetime metric are the largest. For instance, the extreme RN black hole with $q=0$ yields $s_\text{max}=2.1492$ and $r_\text{ISCO}=1.6832$, while the extreme quantum-improved charged black hole with $q=0$ results in $s_\text{max}=2.5274$ and $r_\text{ISCO}=1.8225$. On the other hand, by setting $M=1$, $Q=0.9$, and $q=1$, we obtain $s_\text{max}=2.3761$ and $r_\text{ISCO}=1.4808$ in the RN metric, and $s_\text{max}=2.7605$ and $r_\text{ISCO}=1.5266$ in the vicinity of the quantum-improved charged black hole by setting $M=1$, $Q=0.5$, $\omega=0.5$, and $q=1$. Thus, this means that positively charged particles can attain higher physically possible values of $s_\text{max}$ compared to uncharged particles in the vicinity of black hole extreme cases. However, our numerical analysis may not apply to positively charged particles around extreme cases, where even higher values of $s_\text{max}$ may be achievable. Additionally, if we compare the results of $r_\text{ISCO}$ for charged and uncharged spinning test particles, we can highlight that they are smaller than those reported for uncharged spinless test particles in Table~\ref{table:1} and even smaller than in the extreme cases of the solutions.

    We have observed that the dynamic effects of the test particle's electric charge, $q$, and the spin, $s$, on the ISCO orbit are generally distinct. Both characteristics correct the ISCO parameters differently, depending on the specific black hole solution. An interesting result emerges from considering both characteristics simultaneously, which is evident from the intersection of different curves in Fig.~\ref{fig:fig9}. This outcome implies that test particles with different electric charges $q$ are capable of sharing an orbit if they possess the same spin $s$. The interaction between the spin $s$ and the charge $q$ of the particle could be responsible for this result. As suggested in Ref.~\cite{Zhang2019}, this behavior is observed from the expanded and fully written form of the effective potential in Eq.~\eqref{potentialgeneral}, where various terms involve combinations of factors between $q$ and $s$. However, for a thorough analysis of the problem of multiple charged spinning test particles occupying a single orbit, it is crucial to consider several additional factors, such as the electrical and gravitational interactions between the particles, among others.

\section{Motion of magnetized particles around quantum-improved magnetically charged black holes \label{sec:5}}

    The first consideration of magnetized particles' dynamics around a Schwarzschild black hole in the presence of an external test asymptotically uniform magnetic field is studied in Ref.~\cite{deFelice:2003wx}. Later, studies of magnetized particles' motion around Kerr spacetime, and magnetized and magnetically charged black holes in gravity theories have been developed in Refs.\cite{defelice:2004wz, Abdujabbarov:2020hdp, Rayimbaev:2020gqj, Bokhari:2021qlq, Juraeva:2021gwb, Rayimbaev:2022hrn, Abdulxamidov:2022qwx, Rayimbaev:2022pzr, Rayimbaev:2023vzk, Rayimbaev:2023jmi, Razzaq:2023fdd, Zahid:2022eeq, Nuriddin2022IJMPD}, and it has been found that there is a limit for magnetic interaction parameters in which the ISCO goes to infinity or disappears.

    In the present section, we study magnetized particle motion around quantum-improved magnetically charged black holes. 

%%%%%%%%%%%%%%%%%%%%%%%%%%%%%%%%%%%%%%%%%%%%%%%%%%%%%%%%%%%%

\subsection{Equations of motion} 

    Here, we derive equations of motion of the magnetized particles in the spacetime of the magnetically charged quantum-improved black hole using the Hamilton-Jacobi equation in the form~\cite{deFelice:2003wx}
    \begin{eqnarray}\label{HJ}
    g^{\mu \nu}\frac{\partial {\cal S}}{\partial x^{\mu}} \frac{\partial {\cal S}}{\partial x^{\nu}}=-m^2\Bigg(1-\frac{1}{2m}D^{\mu \nu}F_{\mu \nu} \Bigg)^2\,
    \end{eqnarray}
    where $m$ is the mass of the particle, ${\cal S}$ is the action for the particle, and the term $D^{\mu \nu}F_{\mu \nu}$ stands to describe the interaction between the magnetic field and the dipole moment of the particles. $D^{\mu \nu}$ and $F_{\mu \nu}$ are polarization and electromagnetic field tensors, respectively; $D^{\mu \nu}$ describes the magnetic dipole moment of magnetized particles~\cite{deFelice:2003wx}:
    \begin{eqnarray} \label{Dab}
    D^{\alpha \beta}=\eta^{\alpha \beta \sigma \nu}u_{\sigma}\mu_{\nu} , \qquad D^{\alpha \beta }u_{\beta}=0\ ,
    \end{eqnarray}
    where $\mu^{\nu}$ and $u^{\nu}$ are the four-vector of the dipole moment and the particles measured by the proper observer, respectively. The electromagnetic field tensor $F_{\alpha \beta}$ can be expressed through the electric $E_{\alpha}$ and magnetic $B^{\alpha}$ field components in the following form: 
    \begin{eqnarray}
    F_{\alpha \beta}=u_{[\alpha}E_{\beta]}-\eta_{\alpha \beta \sigma \gamma}u^{\sigma}B^{\gamma}.
    \end{eqnarray}
    
    The term $D^{\mu \nu}$ and $F_{\mu \nu}$ can be calculated by taking into account the condition given in Eq.~(\ref{Dab}) as
    \begin{eqnarray}\label{DF1}
    D^{\mu \nu}F_{\mu \nu}=2\mu^{\hat{\alpha}}B_{\hat{\alpha}}
    =2\mu B_0 \sqrt{f(r)}\ ,
    \end{eqnarray}
    whit $\mu=\sqrt{\mu_{\hat{i}}\mu^{\hat{i}}}$ as the absolute value of the dipole magnetic moment of magnetized particles.

    Now, we investigate the dynamics of magnetized particles around charged improved black holes assuming the black hole is magnetically charged, with the electromagnetic field four-potential,
    \begin{eqnarray}\label{4pot}
    A_{\phi} & = & Q_{m}\cos\theta \ ,
    \end{eqnarray}
    and the non-zero component of the electromagnetic field tensor,
    \begin{eqnarray}\label{FFFF}
     F_{\theta \phi}=-Q_m \sin\theta \ .
    \end{eqnarray}
    
    The orthonormal radial component of the magnetic field generated by the magnetic charge of the improved black hole is given by 
    \begin{equation}\label{BrBt}
    B^{\hat{r}}=\frac{Q_m}{r^2} \ .
    \end{equation}

\subsection{Effective Potential} 

    In this study, we assume the magnetic dipole moment of the magnetized particle parallel to the magnetic field of the black hole that satisfies a stable equilibrium and lies at the equatorial plane, having the components $\mu^{i}=(\mu^{r},0,0)$. In that case, the energy of magnetic interactions between the magnetic dipole of a magnetized particle and the magnetic field of the black hole reaches its minimum. The second part of the condition given in Eq.~\eqref{Dab} allows studying magnetized particles' motion in the proper observer frame. On the other hand, the observer frame helps to avoid the relative motion of the particles and electromagnetic field. We also assume that the absolute value of the magnetic moment has to be constant, and all time it is parallel to the magnetic field of the black hole. Consequently, the interaction term takes the form,
    \begin{eqnarray}\label{DF3} {\cal D}^{\alpha \beta}F_{\alpha \beta} = \frac{2\mu Q_m}{r^2} \ . \end{eqnarray}
    In fact, the axial symmetric properties of the magnetic field of the improved black hole do not change the spacetime symmetries for the magnetized particle motion. So, the two integrals of the motion such as the energy $p_t/m = -e$ and angular momentum $p_{\phi}/m= \ell$ of the particles. 
    We investigate the motion of the magnetized particles in the spacetime of charged improved black hole at the equatorial plane ($\theta=\pi/2$ and $p_{\theta}=0$), using the Hamilton-Jacobi~\eqref{HJ} taking account the scalar product~\eqref{DF1} and we have, 
    \begin{eqnarray}
    \dot{r}^2={e}^2-V_\text{eff}(r)\ .
    \end{eqnarray}
    Then the effective potential for the motion takes the form,
    \begin{eqnarray}\label{effpot}
    V_\text{eff}(r)=f(r) \left[\frac{\ell^2}{r^2}+\left(1-\frac{\cal B}{r^2}\right)^2\right]
    \ ,
    \end{eqnarray}
    where the relation ${\cal B} =\mu Q_m/m$ is the magnetic interaction parameter and $\beta=\mu/(m M)$ is a parameter that characterizes the parameters of the particle and the improved black hole, and it takes only positive values. For the orbital motion of magnetized neutron stars treated as test particles around supermassive or intermediate-mass black holes 
    \begin{eqnarray}
    \beta%=\frac{B_{\rm ns}R_{\rm ns}^3}{2m_{\rm    ns}M_{\rm bh}} 
    \simeq 0.18 \frac{B_{12}R_6^3}{m_1M_6}\ ,
    \end{eqnarray}
    where $B_{12}=B_{\rm ns}/10^{12}\rm G$ normalizes the value of the surface magnetic field of a neutron star to 10$^{12}$ G, while  $R_6=R_{\rm ns}/10^6 \rm cm$ normalizes its radius to 10$^6 cm$. $m_1=m_{\rm ns}/M_{\odot}$ and $M_{\rm bh}/10^6M_{\odot}$ normalizes the masses of neutron stars and black holes to the solar mass $M_{\odot}$, respectively. For example, the parameter $\beta$ for the magnetar SGR (PSR) J1745-2900 ($\mu \simeq 1.6\times 10^{32} \rm G\cdot cm^3$ and  $m \simeq 1.4 M_{\odot}$~\cite{Mori:2013yda}) orbiting the supermassive black hole Sgr A* ($M \simeq 4 \times 10^6M_{\odot}$) is $\beta \simeq 10.2$.

    \begin{figure}[ht!]\centering\includegraphics[width=0.98\linewidth]{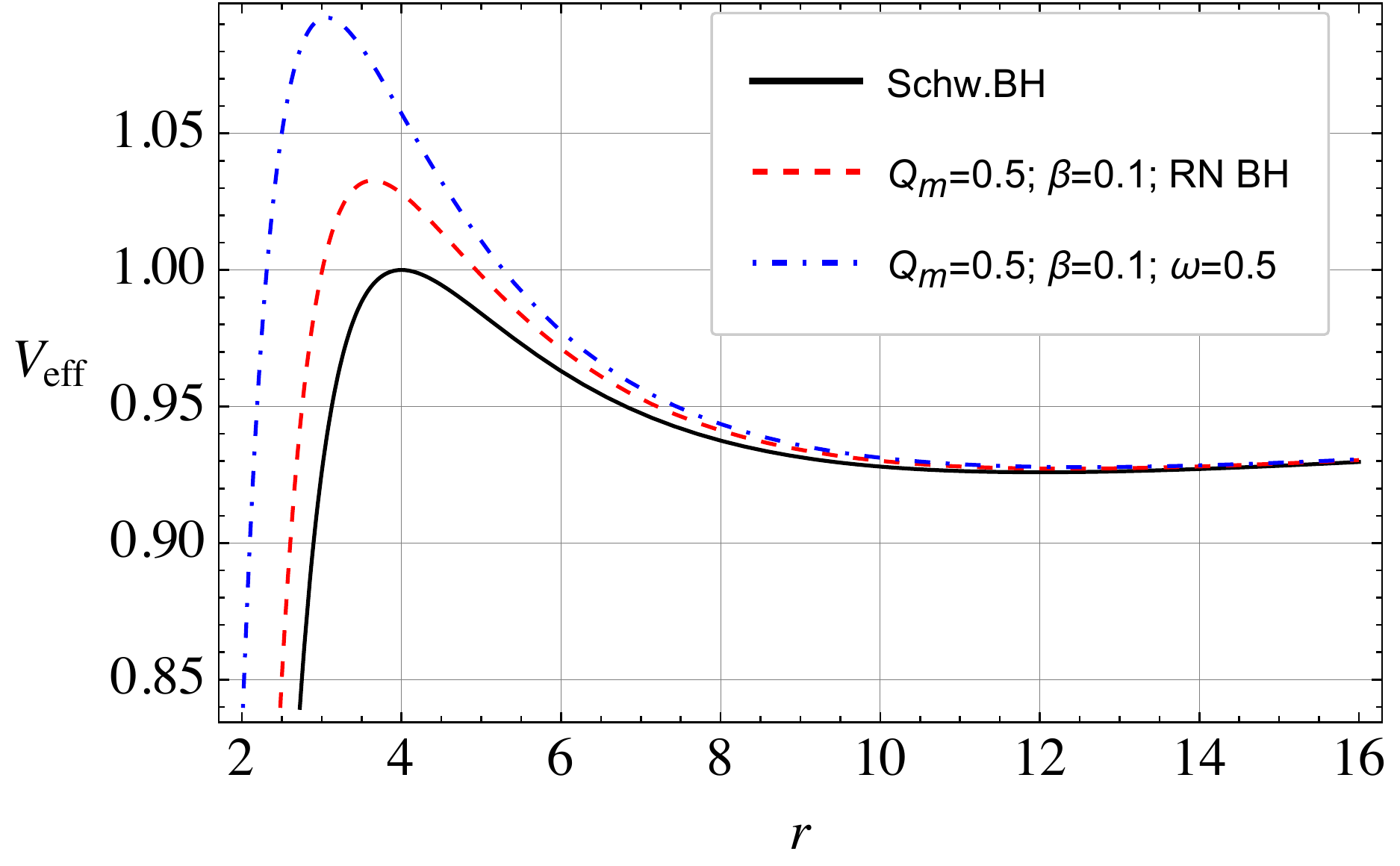}
    \caption{The radial profiles of the effective potential around magnetically charged improved black holes for the different values $Q_m$, $\omega$, and $\beta$. Here, we take $M=1$ and $\ell=4.3$. \label{VV}}\end{figure}
    
    In Fig.~\ref{VV} we present radial profiles of the effective potential for the radial motion of magnetized particles around the magnetically charged quantum-improved black hole. We also compare the profiles with the Schwarzschild and RN black hole cases. It is observed that the presence of the magnetic interaction increases the effective potential as well as the quantum-improved parameter. Moreover, both $\omega$ and $\beta$ parameters shift the orbit position where the effective potential becomes maximum towards the black hole's center.

 \subsection{Circular orbits of test magnetized particles} 
 
    In general, the stability of circular orbits of test particles around a black hole is given by the conditions of Eq.~\eqref{condition}, from which the specific angular momentum and energy of the magnetized particle along the circular orbits can be expressed by the following relations
    \begin{eqnarray}
    \nonumber
    \ell^2&=&\frac{\left(r^2-\beta  M Q_m\right) }{r^2 {\cal Z}(r)}\\\nonumber
    &\times & \Big\{\beta  M Q_m^3 \left(3 r^2+2 \omega \right)+M r^3 \left(r^2-\omega \right)-Q_m^2 r^4\\  &+&\beta  M Q_m \left[2 \left(r^2+\omega \right)^2-M \left(5 r^3+3 r \omega \right)\right]\Big\}
    \ ,
    \\
    {e}^2&=&\frac{\left(r^4-\beta ^2 M^2 Q_m^2\right) \left(-2 M r+Q_m^2+r^2+\omega \right)^2}{r^4 {\cal Z}(r)}, 
    \end{eqnarray}
    where $${\cal Z}(r)=-M r \left(3 r^2+\omega \right)+Q_m^2 \left(2 r^2+\omega \right)+\left(r^2+\omega \right)^2$$

    \begin{figure}[ht!]\centering
    \includegraphics[width=0.98\linewidth]{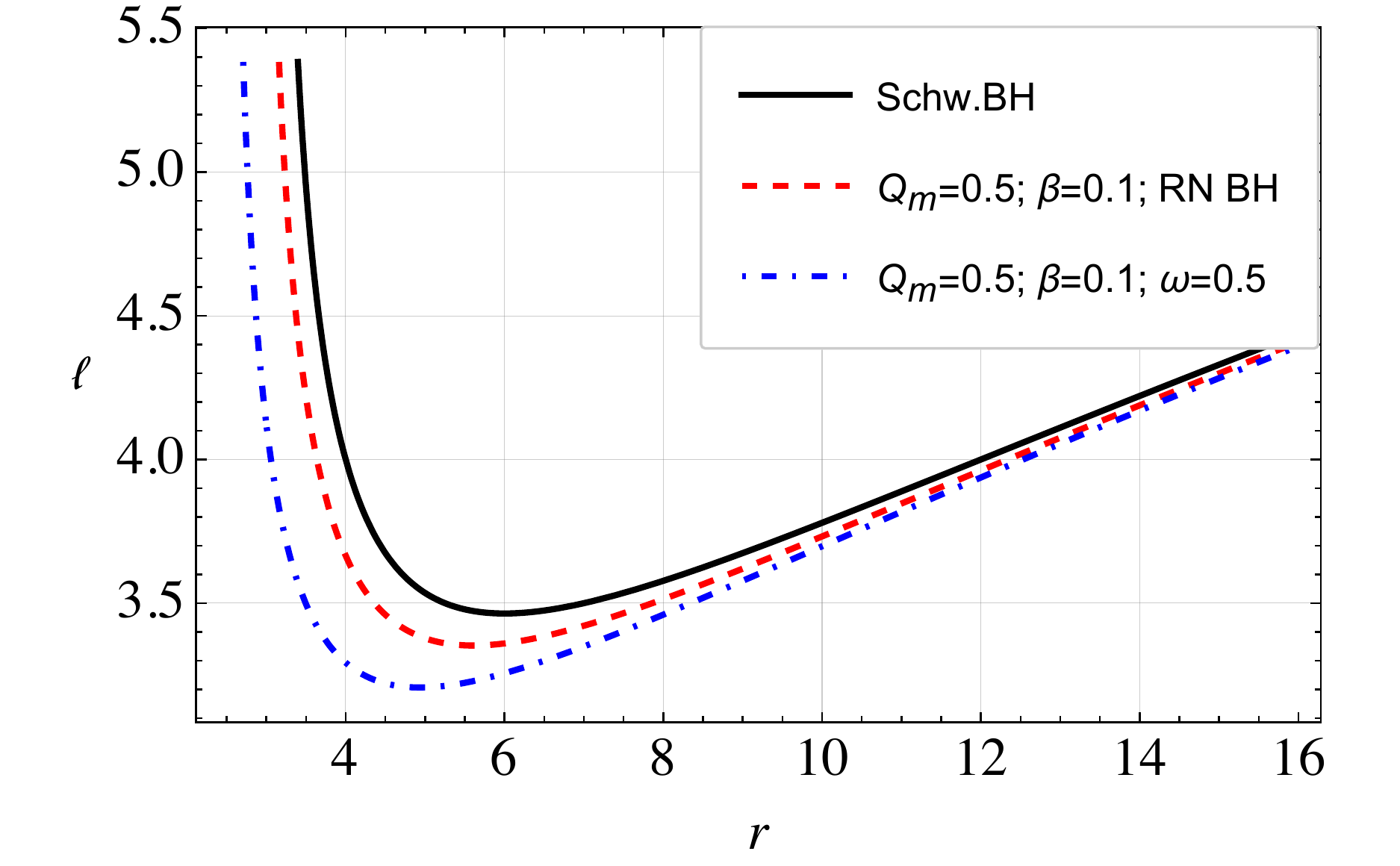}
    \includegraphics[width=0.98\linewidth]{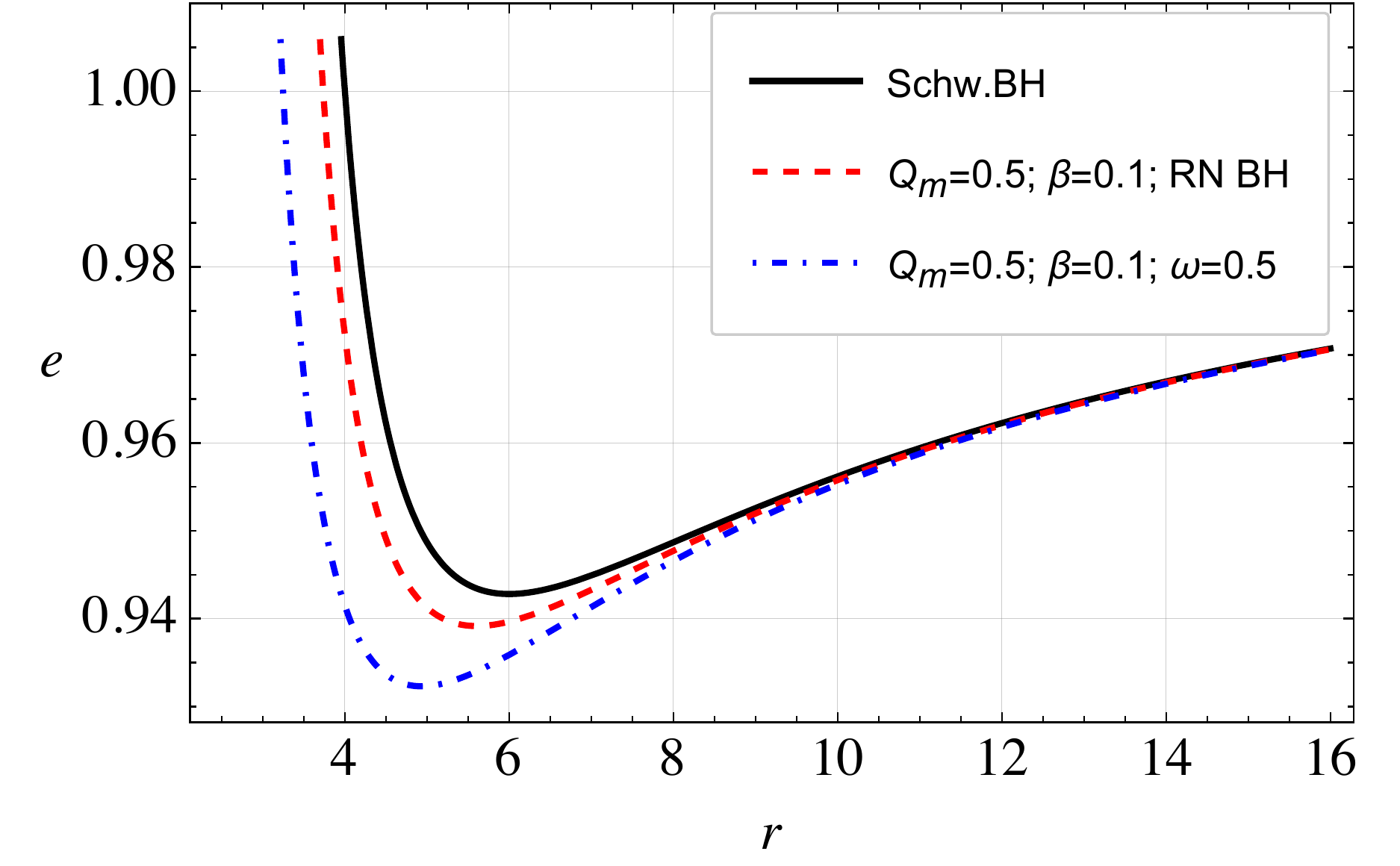}
     \caption{The radial profiles of the specific angular momentum (top panel) and energy (bottom) of magnetized particles corresponding to circular orbits around magnetically charged improved black holes for the different values $Q_m,\omega$, and $\beta$. Here we take as $M=1$. \label{LLgeneric}}
     \end{figure}
    
    Figure~\ref{LLgeneric} demonstrates the radial dependence of the specific angular momentum and energy of magnetized test particles in a circular motion around a magnetically charged improved black hole contrasted with the corresponding circular orbit in Schwarzschild and RN black holes, top and bottom panels, respectively. One can see that the energy and angular momentum decrease in the presence of magnetic interactions and quantum parameters. Moreover, note that the ISCO at $\ell \to \infty$ decreases.

   Observationally speaking, ISCOs around black holes are one of the most important parameters because it corresponds to the inner edge of the accretion disc. One can easily obtain an equation for the ISCO by taking into account the conditions of Eq.~\eqref{condition} for the effective potential of Eq.~\eqref{effpot}, from which one obtains 
    \begin{eqnarray}\label{ISCOEQ}
    \nonumber
    &&Q_m^4 \Big\{8 \beta ^2 M^2 \omega ^2+\beta ^2 M^2 r^3 (12 r-37 M)\\ \nonumber &&+3 \beta ^2 M^2 r \omega  (8 r-3 M)-4 r^6\Big\}+ 4 \beta ^2 M^2 Q_m^6 \left(3 r^2+\omega \right)\\\nonumber &&+Q_m^2 \Big\{6 \beta ^2 M^4 r^2 \left(5 r^2+\omega \right)+M \left(9 r^7-3 r^5 \omega \right)\\ \nonumber &&-4 r^6 \omega -\beta ^2 M^3 r \left(21 r^4+38 r^2 \omega +9 \omega ^2\right)\\ \nonumber &&+4 \beta ^2 M^2 \left(r^2+\omega \right)^3\Big\}+M r^5 \Big\{r^3 (r-6 M)\\&& +2 r \omega  (M+3 r)-3 \omega ^2\Big\}= 0.
    \end{eqnarray}
    It is clear from Eq.~\eqref{ISCOEQ} that in the Schwarzschild limit (when the magnetic charge vanishes $Q_m=0$ and $\omega=0$) the ISCO reduces to $r_{\rm ISCO}=6M$. Since it is difficult to solve analytically Eq.~\eqref{ISCOEQ} with respect to the radial coordinate, we can analyze the ISCO profiles by presenting them in plot form.
    \begin{figure}[ht!]\centering\includegraphics[width=0.9\linewidth]{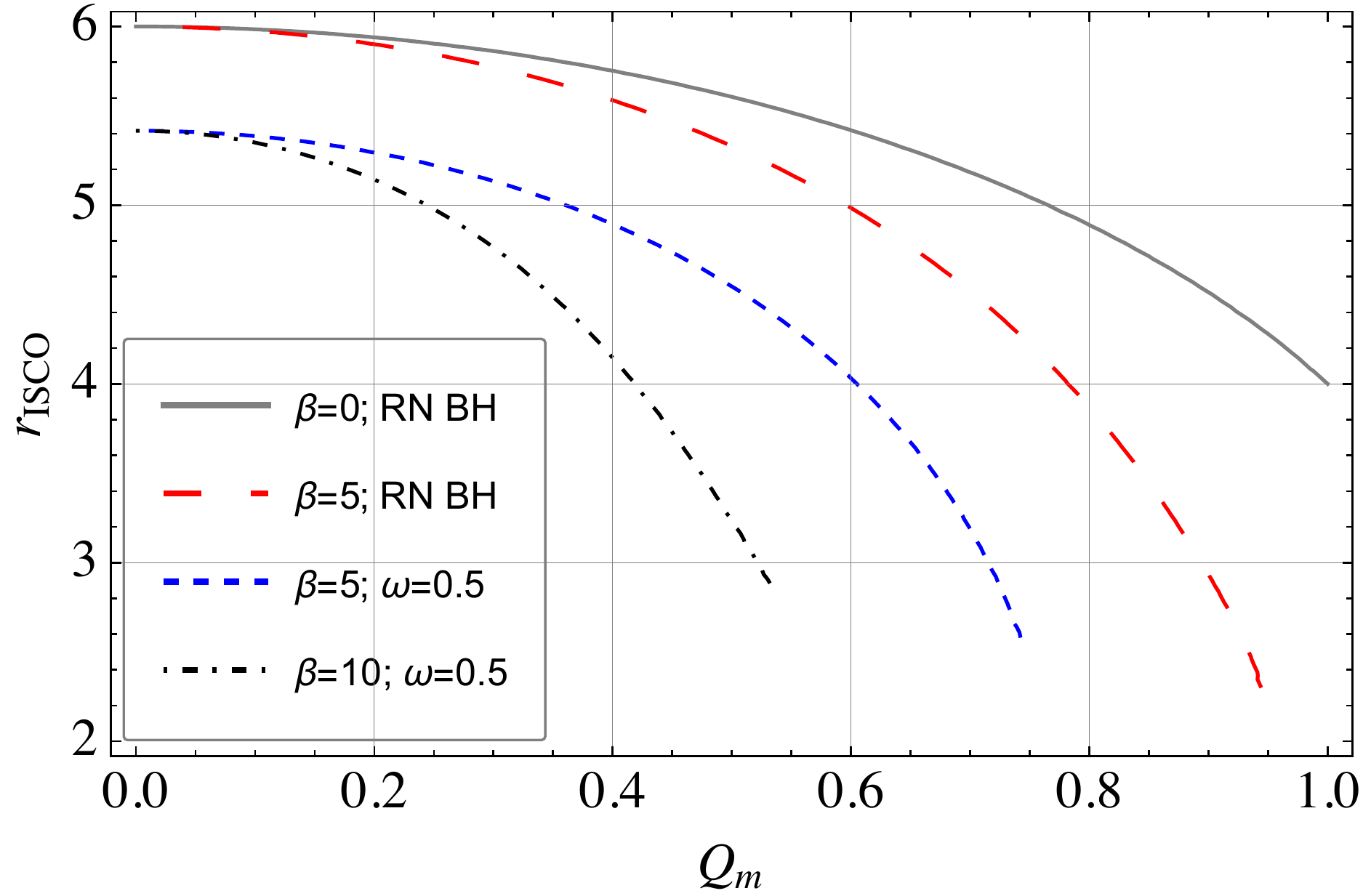}
    \includegraphics[width=0.9\linewidth]{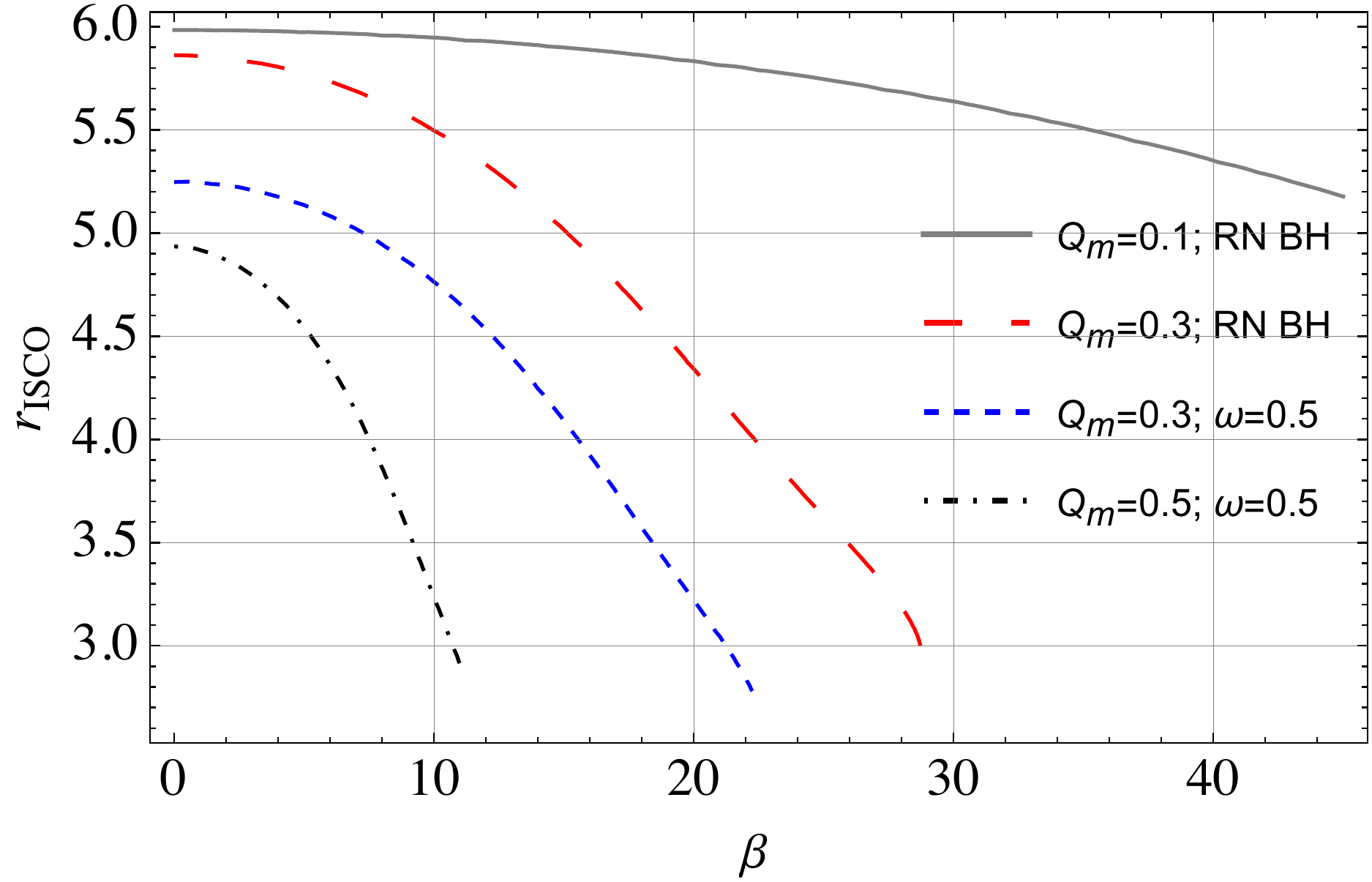}
    \caption{The ISCO radius as a function of the magnetic charge (top panel) and the parameter $\beta$ (bottom). $M=1$. \label{ISCOfig}}\end{figure}

    The effect of the magnetic interaction and quantum corrections on the ISCO for magnetized test particles around a magnetically charged quantum-improved black hole is presented in Fig.~\ref{ISCOfig}. The figure shows how the magnetic charge, $Q_m$, and the parameter $\beta$ reduce the ISCO radius; however, the reduction is faster when $\beta \neq0$. Similarly, an increment in $\omega$ also decreases the ISCO radius for neutral and magnetized test particles. 
    
    Moreover, it is possible to see the existence of an upper limit in the black hole's charge with the presence of $\beta\geq 5$, in which at $Q_m>Q_{up}$ an equilibrium of balanced forces influencing the magnetized test particle will be destroyed, and the particle escapes its ISCO.
    
    Now, we are interested in how $\omega$ and $\beta$ parameters change the upper value in the black hole charge. One can find the value of the charge limit solving Eq.~\eqref{ISCOEQ} numerically in a table form for different values with respect to $r$. Note that the ISCO radius reaches its minimum at this charge limit.

   \begin{figure}[!htb]\centering
    \includegraphics[width=0.9\linewidth]{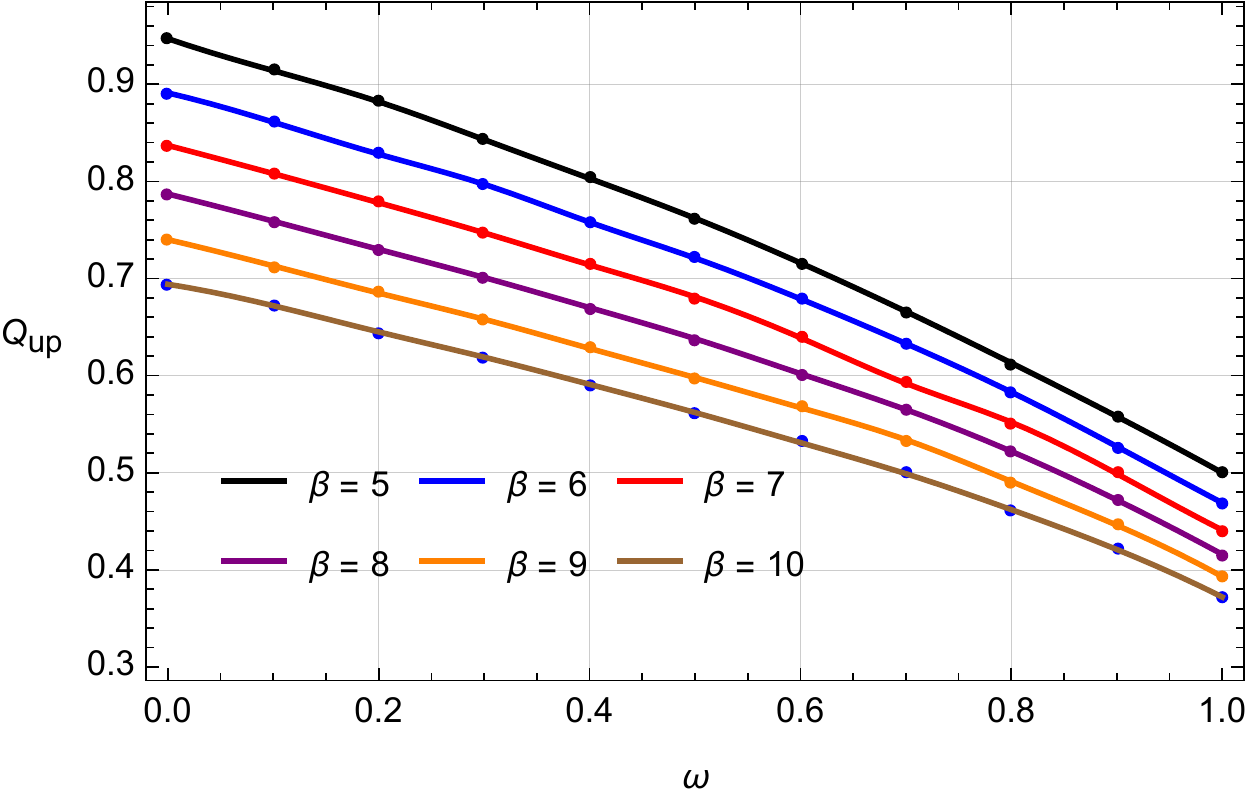}
    \caption{Dependence of the upper limit for the magnetic charge of the black hole $Q_{up}$ as a function of $\omega$ for different values of magnetic interaction parameter.} \label{qvw1}\end{figure}
    
    In Fig.~\ref{qvw1}, we present the dependence of the upper value of the black hole charge corresponding to the existence of ISCOs from the parameter $\omega$ for different values of $\beta$. It is observed from the figure that the upper limit $Q_{up}$ decreases with the increase of both $\omega$ and $\beta$ parameters.

\section{Conclusion} \label{sec:6}

    In this paper, we study the dynamics of charged spinning and magnetized test particles in the spacetime of electrically and magnetically charged quantum-improved black holes, respectively. First, we obtained the relationships between the extreme black hole charge values and the quantum charge parameters, which provide an event horizon in the black hole solution. In deriving the equations of motion of charged spinning test particles, we used the Mathisson-Papapetrou-Dixon equations with an electromagnetic interaction term. In the case of magnetized test particles, on the other hand, we used the Hamilton-Jacobi equation and considered a magnetically charged black hole. 

    We have studied the ISCO parameters, including the radius, angular momentum, and energy, for charged spinless, uncharged spinning, and charged spinning test particles in the charged quantum-improved, Schwarzschild quantum-improved, RN, and Schwarzschild black holes (as well as their extreme cases). Our results reveal that the radius, angular momentum, and energy at the ISCO are smaller for test particles with clockwise spin ($s>0$) and orbit ($j>0$) than for those with anti-clockwise spin ($s<0$). The results also indicate that the ISCO parameters are corrected differently by the electric charge $q$ and spin $s$ of test particles, with the specific black hole solution being a determining factor in the extent of their influence.

    It has been evidenced that the ISCO parameters for charged spinless, uncharged spinning, and charged spinning test particles around a quantum-improved charged black hole are smaller than those for the RN black hole. An increase in $Q$, $\omega$, and $s$ leads to a decrease in the ISCO parameters. Moreover, the quantum correction, represented by $\omega$, has the potential to modify the parabolic trend of $r_\text{ISCO}$ with respect to $q$, causing a decreasing monotonic trend instead. Additionally, higher values of the electric charge $q$ generally correspond to lower values of $\ell_\text{ISCO}$ and higher values of $e_\text{ISCO}$. For a fixed $q$, $r_\text{ISCO}$ decreases monotonically with respect to $s$, but for a given $s$, $r_\text{ISCO}$ does not always exhibit a monotonic behavior. We also saw that in the ISCO orbit of the RN and quantum-improved charged black holes, in the absence of other physical interactions, test particles with different electric charges $q$ can be in the same orbit if they possess the same spin $s$. 

    Our findings also reveal that the quantum parameter $\omega$ increases the maximum spin value $s_\text{max}$, whereas the electric charge $Q$ decreases it. When both parameters are considered simultaneously, $s_\text{max}$ is significantly smaller. Additionally, $s_\text{max}$ can attain larger values as the electric charge $q$ increases. Therefore, the quantum-improved charged black hole has a higher $s_\text{max}$ than the RN black hole, regardless of the value of $q$. As a result, a charged spinning test particle can orbit with higher spin values $s$ if it possesses a more positive electric charge $q$ or larger geometric parameters. Moreover, positively charged spinning particles can achieve higher values of $s_\text{max}$ in the vicinity of black hole extreme cases compared to uncharged spinning particles.

    In the last section, we investigated the magnetized test particle's dynamics around a magnetically charged black hole in quantum-improved QEG. Our analyses show that the presence of $\omega$ causes increasing in the maximum of effective potential and decreasing in the minimum of the energy and angular momentum of the magnetized particles corresponding to circular orbits. Moreover, ISCOs' behavior with respect to $\omega$ has also been studied. It is shown that there is an upper value in the black hole charge, $Q_{up}$, that provides the minimum value for ISCO that a magnetized particle can achieve. Furthermore, we numerically show that the upper limit decreases with the increase of $\beta$ and $\omega$. At $Q_m>Q_\text{up}$ the ISCO does exist due to the dominant effects of the repulsive behavior of the magnetic interaction. Moreover, we found that the upper-value decrease in the presence of $\omega$.

\section*{} \label{sec:7}
\section*{Acknowledgements} \label{sec:acknowledgements}
This work was supported by the Universidad Nacional de Colombia, Hermes Grant Code 57057, and by the Research Incubator No.64 on Computational Astrophysics of the Observatorio Astronómico Nacional. This research is also supported by Grant No. FA-F-2021-510 of the Uzbekistan Agency for Innovative Development. F.A. and J.R. acknowledge the ERASMUS+ ICM project for supporting their stay at the Silesian University in Opava. C.A.B.G. acknowledge the support of the Ministry of Science and Technology of China (grant No.~2020SKA0110201) and the National Science Foundation of China (grants No.~11835009).

\appendix*

\end{document}